\documentclass[aps,prc,twocolumn,showpacs,showkeys,preprintnumbers,footinbib,superscriptaddress,floatfix]{revtex4}
\usepackage{epsfig} 
\usepackage{graphicx}
\usepackage{dcolumn}
\usepackage{xcolor}
\usepackage{amsmath,amssymb}
\usepackage{longtable}
\usepackage{bm}
\usepackage{slashed}
\usepackage{enumerate} 
\usepackage[english]{babel}
\usepackage{textcomp}
\usepackage{hyperref}   
\usepackage{float} 
\usepackage{multirow}
\newcommand{\ld}{lagrangian density}
\newcommand{\mn}{m_{\mathrm{N}}} 
\newcommand{\nb}{n_{\mathrm{B}}} 
\newcommand{\ns}{n_{\sigma}} 
\newcommand{\nr}{n_{\rho}}  
\newcommand{\np}{n_{\mathrm{p}}} 
\newcommand{\nn}{n_{\mathrm{n}}} 
\newcommand{\ism}{isospin symmetric nuclear matter}
\newcommand{\nm}{neutron matter}
\newcommand{\so}{\left(\sigma-\omega\right)}

\newcommand{\La}{\mathcal{L}}
\newcommand{\kf}{\textbf{\textit{k}}_{F}}
\newcommand{\pf}{\textbf{\textit{p}}_{F}}
\newcommand{\kk}{\textbf{\textit{k}}}

\newcommand{\ef}{E_{F}}
\newcommand{\gs}{g_{\sigma}}
\newcommand{\go}{g_{\omega}}
\newcommand{\gr}{g_{\rho}}
\newcommand{\Gs}{G_{\sigma}}
\newcommand{\Go}{G_{\omega}}
\newcommand{\Gr}{G_{\rho}}
\newcommand{\ms}{m_{\sigma}}
\newcommand{\mo}{m_{\omega}}
\newcommand{\mr}{m_{\rho}}
\newcommand{\dd}{\textit{\textbf{d}}}
\begin{document}
\title{Parametrization of the Relativistic ($\sigma-\omega$) Model for Nuclear Matter}
\author{Anis ben Ali Dadi}
\email{Anis.Dadi@mpi-hd.mpg.de}
\affiliation{University of Rostock, Institute of Physics, Universit\"atsplatz 1, 18051 Rostock, Germany}
\affiliation{University of Tunis El Manar, Department of Physics, 1068 Rommana, Tunisia}
\date{\today}
%
%
%
%
\begin{abstract}
\smallskip
\begin{center}
\textbf{Abstract}
\end{center}
We have investigated the zero-temperature equation of state (EoS) for infinite nuclear matter within the $\so$ model at all densities $\nb$ and different proton-neutron asymmetry $\eta\equiv(N-Z)/(N+Z)$. We have presented an analytical expression for the compression modulus, and found that nuclear matter ceases to saturate at $\eta$ slightly larger than $0.8$. Afterward, we have developed an analytical method to determine the strong coupling constants from the EoS for \ism, which allow us to reproduce all the saturation properties with high accuracy. For various values of the nucleon effective mass and the compression modulus, we have found that the quartic self-coupling constant $G_4$ is negative, or positive and very large. Furthermore, we have demonstrated that it is possible (a) to investigate the EoS in terms of $\nb$ and $\eta$; and (b) to reproduce all the known saturation properties without $G_4$. We have thus concluded that the latter is not necessary in the $\so$ model.
\end{abstract}
\maketitle
%
%
%
%
%
%
\section{Introduction}
\label{sec:intro}
Since the beginning of the 20th century, nuclear physics has dealt with the physical properties, structures and reaction mechanisms of the nuclear matter. It is a relatively recent branch of physics, which was developed intensely after the discovery of the atomic nucleus. One of the reasons for this is that its Coulomb field is shielded by the field of the bound electrons in the atom and therefore, it was difficult to prove the existence of the atomic nucleus in our environment. Furthermore, the dimensions of an atomic nucleus are so small that one could describe it technically and from there quantitatively only for the first time in 1911 using Rutherford's scattering experiments. This signalled the birth of nuclear physics.

Nowadays, modern scattering experiments provide direct evidence for the following empirical properties of nuclear matter:
%
%
~(a) The masses of the proton and neutron in the nuclear medium are approximately the same. The force between two protons is equal to the force between two neutrons (i.e., charge independence of the nuclear force), but it depends strongly on the relative spin orientation. These properties can be proven by studying the inelastic scattering of protons and neutrons by deformed nuclei or by studying the mirror nucleus. Therefore, one introduces the concept of isospin symmetry by considering the proton and the neutron as two different quantum states of the same particle, the nucleon.
%
%
~(b) The nuclear force has a very short range (approximately $\pi$-meson Compton wavelength). This property provides a good approximation of the nuclear radius. The estimated value by the scattering of low-energy electrons from nuclei is~\cite{ref:Moeller1988, ref:MyersSwiatecki1998}
\begin{equation}
\label{eq:radius}
r_0\approx1.16~\mathrm{fm},
\end{equation} 
which is $10^5$ times smaller than the atomic radius.
%
~(c) There is no attractive center, which produces the (strong) binding potential. The exact form of the nuclear potential is not precisely known at present, while in the atom, the Coulomb potential ($\propto 1/r$) is responsible for binding the electrons to the nucleus.
%
~(d) A nucleon may only interact (strongly) with a fixed number of its nearest neighbors and next-nearest neighbors. If one adds a further nucleon, only the nuclear volume becomes larger, not the binding energy per nucleon. From this it follows that nuclear matter is a saturated system.
%
~(e) The medium is homogeneous and isotropic; that is, all directions are equivalent and there is no direction preferred to others. This rotational symmetry is, as see in what follows, very helpful
for investigating the equation of state (EoS) for nuclear matter and generally in theoretical nuclear physics.
%
~(f) The mean free path of the nucleon is too much larger than the nucleon size or even larger than the nuclear medium itself, so that one can apply a model of ``independent particles''.

%
%
From these empirical properties, the nuclear medium can be compared with a liquid drop, where nucleons play the role of ``molecules'' in normal liquid. This phenomenological ansatz is known as ``liquid-drop model'', whereby the behavior of the total binding energy $B$ is determined as function of the total number $A$ of nucleons and its different value for different isotopes.
For a given nuclear medium, such as an atomic nucleus of volume $V$ and total mass $M$, consisting of $Z$ bound protons and $N=A-Z$ bound neutrons of free masses $m_{\mathrm{p}}$ and $m_{\mathrm{n}}$, one assumes $B$ to be approximately the sum of different types of energies,
\begin{align}
\label{eq:empirical}
\nonumber
B&=B_{\text{volume}}+B_{\text{asymmetry}}+B_{\text{surface}}+B_{\text{Coulomb}}+...,\\\nonumber
&= a_{\text{v}}\cdot A-a_{\text{sym}}\cdot\frac{(N-Z)^{2}}{A}-a_{\text{surf}}\cdot A^{2/3}-\\
&~~~~a_{\text{C}}\cdot Z(Z-1)A^{-1/3}+...,
\end{align} 
which is equal to the mass defect $(M-Z m_{\mathrm{p}}-N m_{\mathrm{n}})c^2$. This is known as the semi-empirical mass formula~\cite{ref:ABohr}. The coefficients $a_{\text{v}}$, $a_{\text{sym}}$, $a_{\text{surf}}$, and $a_{\text{C}}$ are constants that can be calculated empirically by fitting to experimentally measured masses of different nuclei. In this work we focus only on the first and second terms. The first corresponds to the volume energy, which results from the mutual attraction between nucleons. Because the attraction is short-range and rotationally symmetric, the binding energy per nucleon $E\equiv B/A$ is considered to be independent of $A$, and thus $B$ is proportional to $A$. So the first term gives an explanation of the experimentally observed constant density of nucleons $\nb$, whereby the volume $V$ is proportional to the number $A=V\cdot\nb$ of nucleons. If we simply estimate $V$ as $4\pi r_0^3/3$, using Eq.\eqref{eq:radius}, the saturation density in the nuclear interior amounts an extremely high value of
\begin{equation}
\label{eq:nbsat}
\nb^{sat}=(n_{\text{p}}^{sat}+n_{\text{n}}^{sat})\approx0.153~\mathrm{Nucleon/fm^3},
\end{equation} 
where $\np=Z/V$ and $\nn=N/V$ are the densities of protons and neutrons. Furthermore, the empirical value of the binding energy per nucleon is known with good accuracy from experimental data~\cite{ref:Moeller1988, ref:MyersSwiatecki1998}:
\begin{equation}
\label{eq:energyV}
E_{iso}^{sat}\stackrel{\text{def}}{=}\dfrac{B_{iso}^{sat}}{A}\approx -16.3~\mathrm{MeV}.
\end{equation}
This value can be obtained from the semi-empirical mass formula \eqref{eq:empirical} by considering an isospin symmetric nuclear matter ($\nn=\np$) in the thermodynamic limit ($A,V\to \infty$), so that all terms vanish except the volume energy.\\
The second term in expression \eqref{eq:empirical} has a quantum mechanical meaning, namely, the nuclear medium can be considered as a quantum degenerate Fermi gas, in which all energy levels are occupied separately by protons and neutrons up to the Fermi energy. From the Pauli exclusion principle, each level allows just two particles of the same type with opposite intrinsic spin orientations. Owing to this prevention, the more nucleons are added, the higher are the energy levels that they have to occupy, increasing the total energy of the medium and decreasing the binding energy (the latter will be maximum when protons and neutrons occupy the lowest possible levels). If there is an unequal number of protons and neutrons (usually there are more neutrons than protons), the energies of the highest occupied proton and neutron levels, which are the Fermi energies, are unequal:
$$\dfrac{(\pf^{\text{p}})^2}{2m_{\text{p}}}\neq\dfrac{(\pf^{\text{n}})^2}{2m_{\text{n}}}.$$
In that case, a contribution to the binding energy should assure that protons and neutrons have the same Fermi energy; otherwise, the energy difference would be balanced through $\beta^{\pm}$ decay of neutrons into protons or vice versa. This contribution is the asymmetry energy. Thus, the asymmetry energy provides for an equilibrium between proton and neutron number, vanishes for $\nn=\np$, and lowers the total binding energy $B$ by increasing the difference $\nn-\np$. The empirical value of the asymmetry energy coefficient at the saturation density \eqref{eq:nbsat} is also well known~\cite{ref:Seeger}:
\begin{equation}
\label{eq:asymV}
a^{sat}_{\text{sym}}\approx 32.5~\mathrm{MeV}.
\end{equation}
The next important quantity in the study of nuclear matter properties is the nucleon effective mass $\mn^*$, which results from the mass defect of the nucleon owing to the interaction with its nearest neighbors. From experiments, one determines $\mn^*$ by measuring the density of states of the nucleon and the single-particle energy levels in nuclei~\cite{ref:Brown}. However, a precise value is not known at present. So far as known, the range at saturation density, far away from the Fermi surface and for $\nn=\np$, is~\cite{ref:JaminonMahaux, ref:Johnson}
\begin{equation}
\label{eq:mnstarV}
0.7 \lesssim \mn^*/\mn \lesssim 0.8.
\end{equation}
The compression modulus $\mathcal{K}$ is also an important quantity in the study of nuclear properties, neutron star, supernova collapse and heavy-ion collisions. It is connected with the vibration process in the nucleus, which can be excited to oscillate through inelastic scattering of $\alpha$ particles or through absorption of $\gamma$ quanta. Thereby, there are radial and bending vibrations, in which the deformation of the nucleus changes with the oscillation periods. Concretely, $\mathcal{K}$ can be extrapolated from experimental data on the strength function distribution of the isoscalar giant monopole resonance (GMR) and the isoscalar giant dipole resonance (GDR) in different heavy nuclei~\cite{ref:Blaizot, ref:Youngblood, ref:Garg}. Like the nucleon effective mass, a precise value of the compression modulus is not known at present, but in recent years it has been estimated to be in the range~\cite{ref:Piekarewicz, ref:Garg, ref:Li, ref:Colo} 
\begin{equation}
\label{eq:kV}
230.0 \lesssim \mathcal{K} \lesssim 250.0~\text{MeV}.
\end{equation}
Furthermore, the value $\mathcal{K}\simeq234.0~\text{MeV}$, which we support for the rest of this report, has been estimated by Myers and Swiatecki~\cite{ref:MyersSwiatecki1998} using Thomas-Fermi model calculations. Youngblood \textit{et al.} have reported~\cite{ref:Youngblood} that $\mathcal{K}=231\pm5~\mathrm{MeV}$ by measuring the strength function distribution of the GMR in $^{90}\mathrm{Zr}, ^{116}\mathrm{Sn}, ^{144}\mathrm{Sm}$, and $^{208}\mathrm{Pb}$ using inelastic scattering of $240~\mathrm{MeV}$ $\alpha$ particles at extremely forward angles.

The aim of theoretical nuclear physics is now to develop mathematical models, that provide an accurate description of the observed physical properties in nuclear medium. On one hand, one tries to reproduce the known empirical data as optimally as possible, and on the other hand, one tries to make qualitative and quantitative predictions on the physical behavior of the nuclear medium, which are unknown so far. Because the formulation of the quantum mechanics and the relativity, different theoretical approaches have been developed, which can be divided into two different models:
\begin{enumerate}
\item[i.] Microscopic models based on realistic nucleon-nucleon interaction.
\item[ii.] Effective models based on a phenomenological parametrization of the interaction potentials.
\end{enumerate}
Numerical solution of realistic models, such as the (Dirac-) Br\"uckner-Hartree-Fock theory, is very complicated, so they cannot be applied to finite nuclei. Rather simple and successful approaches generally offer the phenomenological models, such as the Skyrme model, the effective global color model of QCD, the Nambu-Jona-Lasinio model, and the relativistic $\so$ model. The latter is the subject matter of the present report, whose standard version is known as Walecka model~\cite{ref:Walecka} or quantum hadrodynamics (QHD). This effective model provides an accurate description of the nucleon-nucleon interaction at large distances but breaks down at short distances, because it treats baryons and mesons as elementary degrees of freedom and, of course, at short distances quarks and gluons manifest themselves. In contrast, the theory of elementary degrees of freedom of quarks and gluons, that is, QCD, is simple at short distances but complicated at large distances (in the confinement region). The purpose that allows a transition from QCD out into QHD is known as ``hadronization,'' which is, however, not solved so far.

The present article is organized as follows: In Sec~\ref{sec:so}, we calculate the EoS for a static and infinite nuclear matter at finite temperature. Afterward, we focus on the zero-temperature EoS by investigating the case of asymmetric ($\np\neq\nn$), isospin symmetric nuclear ($\nn=\np$) and pure neutron matter ($\np=0$) at all densities. In Sec~\ref{sec:param}, we show how to determine the strong coupling constants analytically from the EoS, which allow us to reproduce the saturation properties of isospin symmetric nuclear matter discussed earlier. That is
\begin{align}
\nonumber
&Input~parameters: \text{Saturation properties},\\
\nonumber
&Output~parameters: \text{Coupling constants},
\end{align} 
which is exactly the opposite of what one usually does. Our conclusions and summary follow in Sec.~\ref{sec:Conclusions}.
We try to use precise values of physical quantities from the recent 2010 CODATA~\cite{ref:codata}; this would be important to minimize the standard deviation errors of the calculated coupling constants from the empirical values and vice versa. Throughout we use the natural Heaviside-Lorentz units $$\hbar=c=k_B=1,$$ where $\hbar$ is the reduced Planck's constant, $c$ is the velocity of light and $k_B$ is the Boltzmann constant. The product of $\hbar$ and $c$ has the dimension of energy times length and the value is~\cite{ref:codata}
$$\hbar\cdot c=197.326950073242~\mathrm{MeV}\cdot\mathrm{fm}=1.$$
The value of the averaged nucleon free mass used is~\cite{ref:codata}
\begin{equation}
\label{eq:mn}
\mn\equiv(m_{\text{p}}+m_{\text{n}})/2=938.91867973~\text{MeV}.
\end{equation}
%
%
%
%
%
%
\section{The ($\sigma-\omega$) Model}
\label{sec:so}
At the saturation density~\eqref{eq:nbsat}, despite the low value of the binding energy~\eqref{eq:energyV} compared with the free mass of the nucleon~\eqref{eq:mn}, a relativistic description is necessarily attributable to the delicate cancellation between short-range repulsion and medium-range attraction. It can also be justified by comparing the Fermi momentum of the nucleon with its free mass, $\kf\approx 0.276\mn$. The $\so$ model has been developed to reproduce the saturation properties of equilibrium nuclear matter and to describe the relativistic effects at higher densities such as in neutron stars~\cite{ref:Glendenning}. The exchange of $\omega$ and $\sigma$ mesons between nucleons is known to provide the short-range repulsion and intermediate-range attraction in the nucleon-nucleon potential. The effective interaction is characterized by the meson parameters such as their masses and coupling constants, which are ``adjusted'' to reproduce the saturation properties.
\begin{figure}[h]
\centering
\scalebox{0.7}{\input{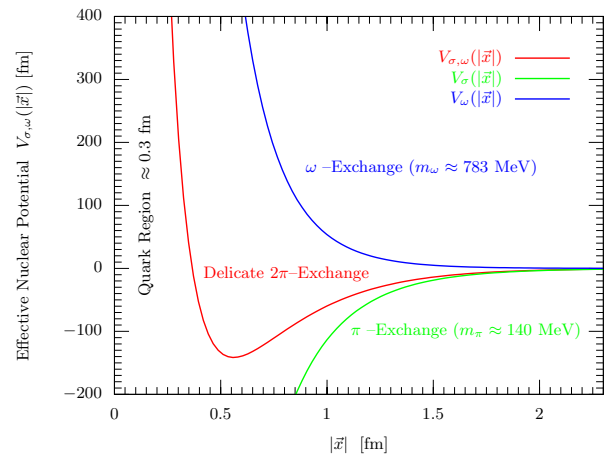}}
\caption{(Color online) Effective nuclear potential in the $(\sigma-\omega)$ model illustrating short-range repulsion and intermediate-range attraction.}
\label{fig:yukawa_fig}
\end{figure}
%
%
This interaction can be expressed phenomenologically for equal numbers of protons and neutrons in terms of two Yukawa potentials,
\begin{align}
\label{eq:yukawa_eq}
\nonumber
V_{\sigma,\omega}(|\vec{x}|)&=V_{\omega}(|\vec{x}|)+V_{\sigma}(|\vec{x}|)\\
&=\left(\dfrac{g_{\omega}^2}{4\pi}\cdot\dfrac{e^{-m_{\omega}|\vec{x}|}}{|\vec{x}|}\right)+\left(-\dfrac{g_{\sigma}^2}{4\pi}\cdot\dfrac{e^{-m_{\sigma}|\vec{x}|}}{|\vec{x}|}\right),
\end{align}
where $\gs$, $\go$, $\ms$ and $\mo$ are the coupling constants and masses of the $\sigma$ and $\omega$ mesons, respectively, and $|\vec{x}|$ is the relative distance between two nucleons. The minus sign tells us that the $\sigma$ meson potential is attractive. Moreover, the heavier (lighter) the exchanged meson is, the more rapidly (slowly) decreases the potential with increasing distance $|\vec{x}|$. Therefore, the range of the resulting interaction~\eqref{eq:yukawa_eq} is mostly determined by the masses of the exchanged $\sigma$ and $\omega$ mesons and the strength by their coupling constants $\gs$ and $\go$ as shown in Fig.~\ref{fig:yukawa_fig}.
%
%
\subsection{The ($\sigma-\omega$) Model at Finite Temperature}
\label{subsec:eosT}
We start our theoretical investigation of nuclear matter properties by defining our system and then constructing the corresponding \ld. We consider a static and uniform system of $A$ bound nucleons ($Z$ protons and $N$ neutrons) moving in a volume $V$, which is assumed to be very large so that we can neglect the surface effects. The total mass and the total binding energy of the system are denoted by $M$ and $B$, respectively. We denote its energy density by $\mathcal{E}\equiv M/V$ and its binding energy per nucleon by $E\equiv B/A=\mathcal{E}/\nb-\mn$, where $\nb$ denotes the density of nucleons, $\nb=(Z+N)/V=\np+\nn$. Each nucleon field depends on space $\vec{x}\equiv(\vec{x}_1, \vec{x}_2, \vec{x}_3)$ and time $t\equiv x^0$ and represented by a Dirac-spinor $\psi(\vec{x},t)=(\mathrm{p},\mathrm{n})^{t}$ as an isospin doublet state of proton and neutron, $\psi=(\mathrm{N}\uparrow,\mathrm{N}\downarrow)^{t}.$\\
The internal structure of nucleons, $\sigma$ and $\omega$ mesons are not taken into account but described as relativistic baryonic and mesonic degrees of freedom. The $\sigma$ meson is isoscalar-scalar and the $\omega$ meson is isoscalar-vector (i.e., both are neutral with total spin $J=0$ and $1$, respectively). As a consequence, they do not have isospin-dependence in the \ld, and therefore they do not contribute in the asymmetry energy. The asymmetry energy originates from the kinetic energy of nucleons (as discussed in the Introduction) and from their interactions itself $(a_{\mathrm{sym}}=a_{\mathrm{kin}}+a_{\mathrm{int}})$. To obtain the empirical value~\eqref{eq:asymV}, we must introduce the $\rho$ meson into the \ld\ of the theory, which is an isovector meson and couples to the isospin density (i.e., to the difference between the density of neutrons and protons, $\nn-\np$). Thus, the \ld\ can be written as
%
%
\begin{eqnarray} 
\label{eq:ld2}
& & \hspace*{-1pc}
\La^{(2)}={\overline\psi}\bigg[\gamma^{\mu}\left(i\partial_{\mu}-\go\chi_{\mu}-\frac{1}{2}\gr\tau b_{\mu}\right) -(\mn-\gs\phi)\bigg]\psi
\nonumber \\ & & \hspace*{1pc}
-\dfrac{1}{2}\ms^2\phi^2+\dfrac{1}{2}\partial_{\mu}\phi\partial^{\mu}\phi+\dfrac{1}{2}\mo^2\chi_{\mu}\chi^{\mu}-\dfrac{1}{4}F_{\mu\nu}F^{\mu\nu}+
\nonumber \\ & & \hspace*{1pc}
\dfrac{1}{2}\mr^2b_{\mu}b^{\mu}-\dfrac{1}{4}B_{\mu\nu}B^{\mu\nu}.
\end{eqnarray}
Here $\tau$ denotes the nucleon isospin matrix; $\phi$ is the field of the scalar $\sigma$ meson; $\chi_{\mu}$ and $F_{\mu\nu}=\partial_{\mu}\chi_{\nu}-\partial_{\nu}\chi_{\mu}$ are the field and strength tensor, respectively, of the vector $\omega$ meson; $b_{\mu}$, $B_{\mu\nu}=\partial_{\mu}b_{\nu}-\partial_{\nu}b_{\mu}$, $\mr$ and $\gr$ are the field, strength tensor, mass, and coupling constant, respectively, of the vector-isovector $\rho$ meson. This \ld\ is invariant under the global $SU(2)$ isospin symmetry and the $U(1)$ gauge symmetry. It contains the Yukawa couplings $\gs\overline\psi\phi\psi$ and $\go\overline\psi\gamma^{\mu}\chi_{\mu}\psi$ of the nucleon field $\psi$ to the scalar and vector meson fields $\phi$ and $\chi_{\mu}$, respectively. The superscript ``(2)'' above $\La$ is related to the quadratic mass term $\ms^2\phi^2$ of the (free) $\sigma$ meson and, as we see later, leading to a linear expression between the scalar $\sigma$ potential and its scalar density. That is why $\La^{(2)}$ is known as the \ld\ of the ``linear $\so$ model''~\cite{ref:Walecka}. In particular, we will see, that this linear version cannot reproduce correctly the nucleon effective mass \eqref{eq:mnstarV} and the compression modulus \eqref{eq:kV}. Therefore, a nonlinear extension of $\La^{(2)}$ introducing scalar self-coupling terms (counterterms) is necessary to reproduce all the five empirical values [Eqs.~\eqref{eq:nbsat}-\eqref{eq:kV}] and to describe the ultrarelativistic limit of the EoS in the high-density approximation (above 0.6 $\mathrm{fm^{-3}}$). Let us then replace the mass term of the free $\sigma$ meson with the following nonlinear scalar potential,
\begin{equation}
\label{eq:nlinear}
\dfrac{1}{2}m_{\sigma}^{2}\phi^{2}\longrightarrow U(\phi)=\textcolor{red}{\dfrac{(\gs\phi)^{2}}{2\Gs}}\textcolor{violet}{+\dfrac{(\gs\phi)^3}{3G_{3}}}\textcolor{blue}{+\dfrac{(\gs\phi)^4}{4G_{4}}},
\end{equation} 
containing, in addition, a three- and four body interaction. The form of this potential is equivalent to that proposed by Boguta and Bodmer~\cite{ref:BogutaBodmer} where 
$$b\equiv\dfrac{1}{\mn G_3}~~\text{and}~~c\equiv\dfrac{1}{G_4}.$$
The \ld\ \eqref{eq:ld2} will be then be written as
%
%
\begin{eqnarray} 
\label{eq:ld4}
& & \hspace*{-1pc}
\La^{(4)}={\overline\psi}\bigg[\gamma^{\mu}\left(i\partial_{\mu}-\go\chi_{\mu}-\frac{1}{2}\gr\tau b_{\mu}\right) -(\mn-\gs\phi)\bigg]\psi
\nonumber \\ & & \hspace*{1pc}
-\textcolor{red}{\dfrac{(\gs\phi)^{2}}{2\Gs}}\textcolor{violet}{-\dfrac{(\gs\phi)^3}{3G_{3}}}\textcolor{blue}{-\dfrac{(\gs\phi)^4}{4G_{4}}}+\dfrac{1}{2}\partial_{\mu}\phi\partial^{\mu}\phi+\dfrac{\go^2}{2\Go}\chi_{\mu}\chi^{\mu}
\nonumber \\ & & \hspace*{1pc}
-\dfrac{1}{4}F_{\mu\nu}F^{\mu\nu}+\dfrac{\gr^2}{2\Gr}b_{\mu}b^{\mu}-\dfrac{1}{4}B_{\mu\nu}B^{\mu\nu}.
\end{eqnarray}
Here we define the following five coupling constants as phenomenological input parameters of the nonlinear model
\begin{equation}
\Gs=\dfrac{\gs^2}{\ms^2},~~\Go=\dfrac{\go^2}{\mo^2},~~\Gr=\dfrac{\gr^2}{\mr^2},~~G_3~~\text{and}~~G_4,
\end{equation} 
with $\dim{\Gs}=\dim{\Go}=\dim{\Gr}=\mathrm{[Mass^{-2}]}$, $\dim{G_3}=\mathrm{[Mass^{-1}]}$ and $G_4$ is dimensionless, because $\dim{\La^{(4)}}\stackrel{!}{\leqslant}\mathrm{[Mass^{4}]}$. From the Hamilton's principle of least action (Euler-Lagrange equations) with respect to $\xi=\{\phi,\chi_{\mu},b_{\mu},\overline\psi\}$
\begin{equation}
\label{eq:hamilton}
\dfrac{\partial \La^{(4)}}{\partial\xi}-\partial_{\mu}\dfrac{\partial \La^{(4)}}{\partial\left(\partial_{\mu}\xi\right)}=0,
\end{equation}
we derive the field equations of the hadronic fields, which are
\begin{subequations}
\label{eq:mvt}
\begin{align}
&\left(\partial_{\mu}\partial^{\mu}+\dfrac{\gs^2}{\Gs}\right)\phi+\dfrac{\gs^3}{G_3}\phi^2+\dfrac{\gs^4}{G_4}\phi^3=\gs\overline\psi\psi, \\
&\partial_{\mu}F^{\mu\nu}+\dfrac{\go^2}{\Go}\chi^{\nu}=\go\overline\psi\gamma^{\nu}\psi, \\
&\partial_{\mu}B^{\mu\nu}+\dfrac{\gr^2}{\Gr}b^{\nu}=\dfrac{1}{2}\gr\overline\psi\tau\gamma^{\nu}\psi, \\
&\left[\gamma^{\mu}\left(i\partial_{\mu}-\go\chi_{\mu}-\frac{1}{2}\gr\tau b_{\mu}\right) -(\mn-\gs\phi)\right]\psi=0. 
\end{align}
\end{subequations}
We now introduce the path integral formalism into statistical physics to calculate the EoS. First, the motion of the nucleon in the volume $V$ is described by the probability amplitude of finding it at the position $\vec{x}'$ at time $t'$, when we know that it was located at point $\vec{x}$ at time $t$. A path integral representation of the probability amplitude can be obtained by discretizing the time interval $t'-t$ up into $L$ infinitesimal segments of fixed duration $\Delta t$ and then taking the continuum limit ($L \to \infty$) at the end of the calculation. Without getting into details, the result is
\begin{align} 
\label{eq:path}\nonumber
&\langle\psi',t'|\psi,t\rangle=\langle\psi|e^{-i\hat{H}^{(4)}(t'-t)}|\psi\rangle\\
&=\int\mathcal{D}\overline\psi \mathcal{D}\psi \mathcal{D}\phi \mathcal{D}\chi_{\mu} \mathcal{D}b_{\mu}\exp \left[\textcolor{red}{i}\int dt\int d^3\vec{x}\La^{(4)}\right],
\end{align}
where $\hat{H}^{(4)}$ is the Hamiltonian of our hadronic system. Second, one of the substantial tools of statistical physics is the grand canonical partition function
\begin{equation}
\label{eq:zt}
\mathcal{Z}(T,V,\mu)=\mathrm{Tr}e^{\textcolor{red}{-}\beta\left(\hat{H}-\mu \hat{A}\right)},
\end{equation}
from which all thermodynamical quantities (energy, pressure, compression modulus, etc.) can be derived. Here $\hat{A}$ is the particle number operator of nucleons and $\mu$ is the nuclear chemical potential. By considering the nuclear medium as a quantum degenerate Fermi gas, the occupied levels can be seen as continuous, because the possible values of the quantized wave number $\kk$ are very close to each other ($\Delta \kk\rightarrow 0$, analog to the electronic band structure). Therefore, one cannot only consider the trace in Eq.~\eqref{eq:zt} as a sum over all microstates, but rather as a phase-space integral; that is,
\begin{equation}
\label{eq:zt_int}
\mathcal{Z}(T,V,\mu)=\int_{\Gamma}\mathcal{D}\psi\langle\psi|e^{-\beta\left(\hat{H}-\mu \hat{A}\right)}|\psi\rangle.
\end{equation}
If we compare now Eq.~\eqref{eq:path} with Eq.~\eqref{eq:zt_int}, we see that the probability amplitude \eqref{eq:path} is expressed in Minkowski space-time, which is not a normed space (i.e., its inner product is not positive-definite). In contrast, the partition function \eqref{eq:zt_int} is defined in phase space $\Gamma$, which is analog to Hilbert space, and this latter is a normed space. So, to unify both expressions, we integrate the \ld\ $\La^{(4)}$ along the imaginary time axis, $t\to -i\tau$, take $\tau=1/T$ as ``inverse temperature,'' and make the substitution $\hat{H}\longrightarrow \hat{H}^{(4)}+\mu\hat{A}$. The partition function \eqref{eq:zt_int} can thus be expressed as a path integral in Euclidean space $(x^4,\vec{x})\equiv(\tau,\vec{x}_1,\vec{x}_2,\vec{x}_3)$ to give
%
%
\begin{eqnarray} 
\label{eq:Z4}
& & \hspace*{-1pc}
\mathcal{Z}\left(\psi,\bar{\psi},\phi,\chi_{\mu},b_{\mu}\right)=\int\mathcal{D}\overline\psi \mathcal{D}\psi \mathcal{D}\phi \mathcal{D}\chi_{\mu} \mathcal{D}b_{\mu}
\nonumber \\ & & \hspace*{1pc}
\exp\left\lbrace \int_0^{\beta}d\tau\int_{V} d^3\vec{x}\left[\La^{(4)}+\mu\overline\psi\gamma^0\psi\right]\right\rbrace.
\end{eqnarray}
To evaluate $\mathcal{Z}$ and to solve the field equations \eqref{eq:mvt}, we use the relativistic mean-field (RMF) approximation under the assumption that the nuclear medium is static and uniformly extended. Notice that this approximation is not directly justified to simplify the calculations considerably, but rather could be seen as a consequence of the observed empirical properties of nuclear matter mentioned in the Introduction. The RMF approximation neglects the fluctuations of the meson fields and replaces them with their mean values, namely,
\begin{subequations}
\label{eq:mean}
\begin{align}
\sigma&\longrightarrow\langle\sigma\rangle=\sigma_0=\gs\phi_0,\\
\omega_{\mu}&\longrightarrow\langle\omega_{\mu}\rangle=\omega_{0}=\go\chi_0,\\
\rho_{\mu}&\longrightarrow\langle\rho_{\mu}\rangle=\rho_{0}=\gr b_0,
\end{align}
\end{subequations}
which means that protons and neutrons move independently in classical mean fields $\phi_0$, $\chi_0$, and $b_0$. As a first consequence, the field equations \eqref{eq:mvt} simplify considerably to give
\begin{subequations}
\label{eq:mvt0}
\begin{align}
\label{eq:mvts0}
&\sigma_0+\dfrac{\Gs}{G_3}\sigma_0^2+\dfrac{\Gs}{G_4}\sigma_0^3=\Gs\langle\overline\psi\psi\rangle\stackrel{\text{def}}{=}\Gs \ns, \\
\label{eq:mvto0}
&\omega_0=\Go\langle\overline\psi\gamma^0\psi\rangle=\Go \nb= \Go\left( n_{\mathrm{p}}+n_{\mathrm{n}}\right), \\
\label{eq:mvtr0}
&\rho_0=\dfrac{1}{2}\Gr\langle\overline\psi\gamma^0\tau_3\psi\rangle=\dfrac{1}{2}\Gr\left( n_{\mathrm{p}}-n_{\mathrm{n}}\right),\\
\label{eq:mvtpsi0}
&\left[i\gamma_{\mu}\partial^{\mu}-\gamma_{0}\left(\omega_0+\dfrac{1}{2}\tau_3\rho_0\right)-(\mn-\sigma_0)\right]\psi(\vec{x},t)=0.
\end{align}
\end{subequations}
In Eq.~\eqref{eq:mvts0} we defined a Lorentz scalar density $\ns$ of the nucleons, which depends on the Fermi momenta of protons and neutrons, $\kf^{\text{p}}$ and $\kf^{\text{n}}$. In Eq.~\eqref{eq:mvtr0} we have $\tau_3=+1$ for protons and $\tau_3=-1$ for neutrons. Further, the partition function \eqref{eq:Z4} simply yields
%
%
\begin{eqnarray} 
\label{eq:Z_RMF4}
& & \hspace*{0pc}
\mathcal{Z}\left(\sigma_0,\omega_0,\rho_0\right)=const\times
\nonumber \\ & & \hspace*{0pc}
\exp\Bigg[\beta V\left(\dfrac{\omega_0^2}{2\Go}+\dfrac{\rho_0^2}{2\Gr}-\dfrac{ \sigma_0^2}{2\Gs}-\dfrac{\sigma_0^3}{3G_3}-\dfrac{\sigma_0^4}{4G_4}\right)
\nonumber \\ & & \hspace*{0pc}
+\dfrac{\dd\cdot V}{2\pi^2}\int_0^{\kf}\kk^2d\kk\bigg[\ln{\left(1+e^{-\beta\left(E_0^*-\mu_0^*\right)}\right)}
\nonumber \\ & & \hspace*{0pc}
+\ln{\left(1+e^{-\beta\left(E_0^*+\mu_0^*\right)}\right)}\bigg]\Bigg],
\end{eqnarray}
%
where we define, respectively,
\begin{align}
\label{eq:estardef}
E_0^*&\stackrel{\text{def}}{=}\sqrt{\kk^2+\mn^{*2}}, \\
\label{eq:mnstardef}
\mn^*&\stackrel{\text{def}}{=}\mn-\sigma_0, \\
\label{eq:mustardef}
\mu_0^*&\stackrel{\text{def}}{=}\mu-\omega_0-\dfrac{1}{2}\tau_3\rho_0,
\end{align}
as an effective dispersion relation, effective mass, and effective chemical potential (or effective Fermi energy, $\ef^*$) of the nucleon. The $\dd$ in Eq.~\eqref{eq:Z_RMF4} denotes the degeneracy factor, having the value 4 for isospin symmetric nuclear matter ($\nn=\np$) and 2 for pure neutron matter ($\np=0$). Rigorously speaking, the partition function \eqref{eq:Z_RMF4} must be expressed in terms of product over $\kf^{\text{p}}$ and $\kf^{\text{n}}$, and then taking $\dd=2$. From Eq.~\eqref{eq:mustardef} it becomes clear that the isospin forces arising from the $\rho$ meson exchange lead to add the quantity $\rho_0/2$ to the neutron Fermi energy and to substract the same quantity from the proton Fermi energy, while the repulsive forces arising from the $\omega$ meson exchange lead to decrease the Fermi energy of the proton from $(\mu^{\text{p}}-\rho_0/2)$ to $(\mu^{\text{p}}-\rho_0/2)-\omega_0$ and of the neutron from $(\mu^{\text{n}}+\rho_0/2)$ to $(\mu^{\text{n}}+\rho_0/2)-\omega_0$. Moreover, the attractive forces arising from the $\sigma$ meson exchange lead to decrease the nucleon mass from $\mn$ to $\mn^*$ and this mass defect becomes more important by introducing the three- and four-body interactions \eqref{eq:nlinear}. The scalar mean fields $\sigma_0$, $\omega_0$, and $\rho_0$ have the dimensions of mass; the mean effective quantities $\mn^*$ and $\mu_0^*$ are scalar in Dirac space and depend on the Fermi momenta of protons and neutrons.\\

Now we are able to derive the EoS for infinite nuclear matter and to investigate its thermodynamics at given values of $N$ and $Z$. In thermodynamic equilibrium, the meson-exchange process is reduced substantially, leading to a maximum entropy $S$, which is
\begin{equation}
\label{eq:eqcond}
\dfrac{\partial S}{\partial \sigma_0}=\dfrac{\partial S}{\partial \omega_0}=\dfrac{\partial S}{\partial \rho_0}\stackrel{\text{!}}{=}0,
\end{equation}
with $S=\ln{\mathcal{Z}(\sigma_0,\omega_0,\rho_0)}$. So, using Eq.~\eqref{eq:Z_RMF4}, making the derivative of $S$ with respect to $\sigma_0$ and identifying the result with Eq.~\eqref{eq:mvts0}, we obtain the scalar densities of protons and neutrons at a given temperature $T$
\begin{subequations}
\label{eq:dpn}
%
%
\begin{eqnarray} 
\label{eq:scalardp}
& & \hspace*{-1pc}
\ns^{\text{p}}\left(\kf^{\text{p}},\mu^{\text{p}},T\right)=
\\ \nonumber & & \hspace*{-1pc}
\dfrac{1}{\pi^2}\int_0^{\kf^{\text{p}}}\kk^2d\kk\dfrac{\mn^*}{E_0^*}\cdot\bigg(\dfrac{1}{e^{\beta\left(E_0^*-\mu_0^*\right)}+1}+\dfrac{1}{e^{\beta\left(E_0^*+\mu_0^*\right)}+1}\bigg),\\
\label{eq:scalardn}
& & \hspace*{-1pc}
\ns^{\text{n}}\left(\kf^{\text{n}},\mu^{\text{n}},T\right)=
\\ \nonumber & & \hspace*{-1pc}
\dfrac{1}{\pi^2}\int_0^{\kf^{\text{n}}}\kk^2d\kk\dfrac{\mn^*}{E_0^*}\cdot\bigg(\dfrac{1}{e^{\beta\left(E_0^*-\mu_0^*\right)}+1}+\dfrac{1}{e^{\beta\left(E_0^*+\mu_0^*\right)}+1}\bigg),
\end{eqnarray}
with $\ns^{\text{p}}+\ns^{\text{n}}=\ns$~[Eq.~\eqref{eq:mvts0}]. On the right-hand side of Eq.~\eqref{eq:scalardp}, we see that the scalar density of protons is expressed as a sum of two thermal distribution functions for a relativistic Fermi gas of protons and antiprotons. The first has an effective mass $\mn^*$ and chemical potential $\mu^*_0$, whereas the second one has the same effective mass but opposite chemical potential, $-\mu^*_0$, so that the total proton number remains constant (from the beginning, we do not consider for example the $\beta$ decay). Further, the sum of both distributions is divided by the Lorentz factor $E_0^*/\mn^*$, which is an effect of the Lorentz contraction; the higher the density of protons is, the more important this relativistic effect will be. These interpretations are analogous to that for the scalar density of neutrons in Eq.~\eqref{eq:scalardn}.\\
Similarly, the partial derivative of $S$ with respect to $\omega_0$ and $\rho_0$ yields the ``net'' baryon densities of protons and neutrons,
%
%
\begin{eqnarray} 
\label{eq:protond}
& & \hspace*{-1pc}
n_{\text{p}}\left(\kf^{\text{p}},\mu^{\text{p}},T\right)=
\\ \nonumber & & \hspace*{-1pc}
\dfrac{1}{\pi^2}\int_0^{\kf^{\text{p}}}\kk^2d\kk\cdot\bigg(\dfrac{1}{e^{\beta\left(E_0^*-\mu_0^*\right)}+1}-\dfrac{1}{e^{\beta\left(E_0^*+\mu_0^*\right)}+1}\bigg),\\
\label{eq:neutrond}
& & \hspace*{-1pc}
n_{\text{n}}\left(\kf^{\text{n}},\mu^{\text{n}},T\right)=
\\ \nonumber & & \hspace*{-1pc}
\dfrac{1}{\pi^2}\int_0^{\kf^{\text{n}}}\kk^2d\kk\cdot\bigg(\dfrac{1}{e^{\beta\left(E_0^*-\mu_0^*\right)}+1}-\dfrac{1}{e^{\beta\left(E_0^*+\mu_0^*\right)}+1}\bigg).
\end{eqnarray}
\end{subequations}
%

%
\subsection{Equation of State at Zero Temperature}
\label{subsec:eosT0}
At zero temperature, the thermal Fermi distributions of baryons in Eqs.\eqref{eq:dpn} give one, and of antibaryons give zero, because we still have $\kk\leqslant \kf\Rightarrow E_0^*\leqslant \ef^*\equiv\mu_0^*$. Thus, Eqs. \eqref{eq:protond} and \eqref{eq:neutrond} yield the well known expressions for the baryon density,
\begin{subequations}
\label{eq:d0}
\begin{align}
\label{eq:bd0}
\nb&=\frac{1}{3\pi^2}\left[\left(\kf^{\text{p}}\right)^3+\left( \kf^{\text{n}}\right)^3\right],
\end{align}
and the isospin density,
\begin{align}
\label{eq:isod0}
\nr&=\frac{1}{3\pi^2}\left[\left(\kf^{\text{p}}\right)^3-\left( \kf^{\text{n}}\right)^3\right].
\end{align}
\end{subequations}
As an application, for isospin symmetric nuclear matter at saturation density \eqref{eq:nbsat}, we obtain
\begin{equation}
\label{eq:pfiso}
\kf^{\text{p}}=\kf^{\text{n}}\approx259.147~\mathrm{MeV}\approx1.313~\mathrm{fm^{-1}}.
\end{equation} 
Now using Eqs.~\eqref{eq:mnstardef}, \eqref{eq:mvts0}, and \eqref{eq:dpn}, we deduce the first zero-temperature EoS at given values of proton and neutron densities, namely, the nucleon effective mass
%
%
\begin{align} 
\label{eq:mnstarInt}
\nonumber
\mn^*=&~\mn+\dfrac{\Gs}{G_3}\sigma_0^2+\dfrac{\Gs}{G_4}\sigma_0^3- \\
&\dfrac{\Gs}{\pi^2}\left[\int_0^{\kf^{\text{p}}}\kk^2d\kk\dfrac{\mn^*}{E_0^*}+\int_0^{\kf^{\text{n}}}\kk^2d\kk\dfrac{\mn^*}{E_0^*}\right].
\end{align}
The pressure can be deduced from the grand canonical potential $\mathcal{J}=-T\cdot\ln{\mathcal{Z}}(T,V,\mu)=-\mathcal{P}\cdot V$ to give
%
%
\begin{align} 
\label{eq:pressureInt}
\nonumber
&\mathcal{P}\left(\np,\nn\right)=\dfrac{1}{2}\Go\left(\np+\nn\right)^2+\dfrac{1}{8}\Gr\left(\nn-\np\right)^2-\frac{\sigma_0^2}{2 \Gs}-\\ 
&\frac{\sigma_0^3}{3 G_3}-\frac{\sigma_0^4}{4G_4}+ \dfrac{1}{3\pi^2}\left[\int_0^{\kf^{\text{p}}}\dfrac{\kk^4d\kk}{E_0^*}+\int_0^{\kf^{\text{n}}}\dfrac{\kk^4d\kk}{E_0^*}\right].
\end{align}
Further, the energy density $\mathcal{E}=M/V$ can be deduced from the Gibbs-Duhem equation, which is simply given by
\begin{eqnarray}
\label{eq:gibbs}
\mathcal{E}(\nb)=\mu\cdot n_B-\mathcal{P}(\nb),
\end{eqnarray}
at zero temperature. This, together with Eqs.~\eqref{eq:mvto0} and \eqref{eq:mustardef} yields
%
%
%
\begin{align} 
\label{eq:energydInt}
\nonumber
&\mathcal{E}\left(\np,\nn\right)=\dfrac{1}{2}\Go\left(\np+\nn\right)^2+\dfrac{1}{8}\Gr\left(\nn-\np\right)^2+\frac{\sigma_0^2}{2 \Gs}+\\
&\frac{\sigma_0^3}{3 G_3}+\frac{\sigma_0^4}{4G_4}+ \dfrac{1}{\pi^2}\left[\int_0^{\kf^{\text{p}}}d\kk\kk^2E_0^*+\int_0^{\kf^{\text{n}}}d\kk\kk^2E_0^*\right].
\end{align}
Notice that all integrals over the Fermi momenta appearing in Eqs.~\eqref{eq:mnstarInt}, \eqref{eq:pressureInt}, and \eqref{eq:energydInt} can be done analytically, and then they can be expressed solely in terms of $\kf^{\text{p}}$ and $\kf^{\text{n}}$ or, according to Eq.~\eqref{eq:d0}, the baryon densities $\np$ and $\nn$. Further, from the semi-empirical mass formula \eqref{eq:empirical}, we see that the asymmetry energy is given by
\begin{equation}
\label{eq:asymstart}
a_{\text{sym}}=\dfrac{1}{2}\Big|\left(\dfrac{\partial^2E}{\partial\eta^2}\right)_{\nb \text{fixed}}\Big|_{\eta=0},
\end{equation}
where $\eta\equiv\nr/\nb=(\nn-\np)/(\nn+\np)$ denotes the proton-neutron fraction. To calculate $a_{\text{sym}}$ within the $\so$ model, it would be easier to express $\np$ and $\nn$ in term of $\eta$ and $\nb$,
\begin{subequations}
\label{eq:npnneta}
\begin{align}
\np&=\dfrac{\nb}{2}(1-\eta),\\
\nn&=\dfrac{\nb}{2}(1+\eta),
\end{align}
\end{subequations}
and then make the second derivative of the energy density \eqref{eq:energydInt} with respect to $\eta$. After some calculation, we obtain
\begin{equation}
\label{eq:asymeq}
a_{\text{sym}}=\dfrac{\kf^2}{6\ef^*}+\dfrac{\Gr}{8}\nb=a_{\mathrm{kin}}+a_{\mathrm{int}}.
\end{equation} 
The first term $a_{\mathrm{kin}}$ arises from the difference between the Fermi energies of protons and neutrons (see Introduction). The second $a_{\mathrm{int}}$ arises from the isospin coupling between the $\rho$ mesons and the nucleons. Both together, $a_{\mathrm{kin}}+a_{\mathrm{int}}$, assure that neutrons and protons have the same Fermi energy.\\
Alternatively, we can easily prove from Eq.~\eqref{eq:empirical}, that $a_{\text{sym}}$ is the difference between the binding energy per nucleon $E(\np, \nn)$ of pure \nm\ and that of \ism. Namely, from Eq.~\eqref{eq:npnneta} we have $\eta=1$ for pure neutron matter, $\eta=0$ for isospin symmetric nuclear matter and $0<\eta<1$ for asymmetric nuclear matter ($Z\neq N$). So, if we write the binding energy per nucleon at given $Z$ and $N$ as
$$E(\nb,\eta)=E_{iso}(\nb)+\eta^2\cdot a_{\text{sym}}+\mathcal{O}(\eta^4),$$
then we get
\begin{equation}
\label{eq:asymeq2}
a_{\text{sym}}=E(\nb,1)-E(\nb,0).
\end{equation} 
\\We can thus compare the numerical result of Eq.~\eqref{eq:asymeq2} with that of Eq.~\eqref{eq:asymeq}. In addition, we can investigate the evolution of the energy density \eqref{eq:energydInt} at all values of the proton-neutron fraction $\eta$ between $0$ and $1$, which now depends solely on this parameter and $\nb$.

We close the theoretical framework by calculating the compression modulus. By starting from its definition and using the first law of thermodynamics, $dB=-PdV$, we obtain
\begin{eqnarray}
\label{eq:Kdef}
\mathcal{K}(\eta)\stackrel{\text{def}}{=}\kf^2\dfrac{\partial^2E}{\partial\kf^2}=9\cdot\left(\dfrac{\partial\mathcal{P}\left(\eta,\nb\right)}{\partial\nb}\right)_{\nb=\nb^{sat}, T=0}.
\end{eqnarray}
\\This simply means that the compression modulus is the slope of the pressure \eqref{eq:pressureInt} at saturation density and multiplied by $9$. Furthermore, $\mathcal{K}$ may be expressed analytically in terms of $\eta$ and the saturation density by evaluating the derivative of the scalar mean field $\sigma_0$ in Eq.~\eqref{eq:mvts0} with respect to $\nb$, or equivalently to $\kf$, which yields the following expression in terms of $\eta$ and $\nb$:
\begin{align}
\label{eq:dseta}
\dfrac{d\sigma_0}{d\nb}(\nb, \eta)=\dfrac{X_1}{2\Gs^{-1}+X_2-X_3},
\end{align}
with
\begin{align}
\nonumber
&X_1=\mn^*\bigg[(\ef^{*\text{n}})^{-1}+(\ef^{*\text{p}})^{-1}+\eta\bigg((\ef^{*\text{n}})^{-1}-(\ef^{*\text{p}})^{-1}\bigg)\bigg],\\\nonumber
&X_2=\left[\dfrac{\kf^{\text{n}3}+3\kf^{\text{n}}\mn^{*2}}{\pi^2\ef^{*\text{n}}}+\dfrac{\kf^{\text{p}3}+3\kf^{\text{p}}\mn^{*2}}{\pi^2\ef^{*\text{p}}}+\dfrac{4\sigma_0}{G_3}+\dfrac{6\sigma_0^2}{G_4}\right],\\\nonumber
&X_3=-\dfrac{3}{\pi^2}\mn^{*2}\left[\ln\left(\dfrac{\kf^{\text{n}}+\ef^{*\text{n}}}{\mn^*}\right)+\ln\left(\dfrac{\kf^{\text{p}}+\ef^{*\text{p}}}{\mn^*}\right)\right],\\\nonumber
&\ef^{*\text{p}}=\sqrt{\kf^{\text{p}2}+\mn^{*2}},\\\nonumber
&\ef^{*\text{n}}=\sqrt{\kf^{\text{n}2}+\mn^{*2}},\\\nonumber
&\kf^{\text{p}}=(3\pi^2\np)^{1/3},\\\nonumber
&\kf^{\text{n}}=(3\pi^2\nn)^{1/3}.
\end{align}
This, together with Eq.~\eqref{eq:Kdef} and the expression of the pressure \eqref{eq:pressureInt} yields $\mathcal{K}$ for cold nuclear matter in terms of $\eta$ and $\nb^{sat}$
\begin{widetext}
%
%
\begin{align}
\label{eq:Ketaeq}
\nonumber
&\mathcal{K}(\nb^{sat},\eta)=9\left(\Go+\dfrac{1}{4}\Gr\eta^2\right)\nb-9\left(\dfrac{\sigma_0}{\Gs}+\dfrac{\sigma_0^2}{G_3}+\dfrac{\sigma_0^3}{G_4}\right)\dfrac{d\sigma_0}{d\nb}+\dfrac{3}{2}\left(\dfrac{\kf^{\text{n}2}}{\ef^{*\text{n}}}-\dfrac{\kf^{\text{p}2}}{\ef^{*\text{p}}}\right)\eta+\sum_{\kf=\kf^{\text{p}},\kf^{\text{n}}}\left[\dfrac{3\kf^2}{2\ef^*}\right]+\\
&\dfrac{d\sigma_0}{d\nb}\times\sum_{\kf=\kf^{\text{p}},\kf^{\text{n}}}\left[\dfrac{3\mn^*\kf^3}{2\pi^2\ef^*}+\dfrac{9\kf\mn^{*3}}{2\pi^2\ef^*}-\dfrac{9}{2\pi^2}\mn^{*3}\ln\left(\dfrac{\ef^*+\kf}{\mn^*}\right)\right].
\end{align}
\end{widetext}
We point out that all quantities appearing in Eq.~\eqref{eq:Ketaeq}, as well as the derivative of the scalar mean field \eqref{eq:dseta}, should be evaluated at $\nb=\nb^{sat}$. 
%
\subsection{Results}
\label{subsec:results}
Before discussing the results, we should make clear the meaning of the following notations:
\begin{itemize}
\item \textcolor{red}{RMF2} means the relativistic mean-field approximation to the linear $\so$ model, that is, without three- and four-body self-interactions. The \ld\ used for this is given by Eq.~\eqref{eq:ld2}. There are thus three coupling constants: $\Gs$, $\Go$, and $\Gr$.
\item \textcolor{violet}{RMF3} means the relativistic mean-field approximation to the nonlinear $\so$ model using the \ld\ \eqref{eq:ld4} but without the four-body self-interaction. The coupling constants are four: $\Gs$, $\Go$, $\Gr$, and $G_3$.
\item \textcolor{blue}{RMF4} means the relativistic mean-field approximation to the nonlinear $\so$ model by introducing the three- and four-body self-interactions. The coupling constants are five: $\Gs$, $\Go$, $\Gr$, $G_3$, and $G_4$.
\end{itemize}
One of our aims is to reproduce the five saturation properties of \ism\ for each model by determining separately the three, four, and five coupling constants of RMF2, RMF3, and RMF4, respectively. Notice that only Eq.~\eqref{eq:mnstarInt} must be solved numerically, that is,
$$\mn^*\stackrel{\text{?}}{=}f(\nb, \mn^*),$$
and the other EoS \eqref{eq:pressureInt}, \eqref{eq:energydInt}, \eqref{eq:asymeq}, and \eqref{eq:Ketaeq} can be deduced analytically. For example, using a Fortran-95 code for the secant method~\cite{ref:Pang}, the error we obtained is roughly $10^{-8}$ comparing with an exact solution. Throughout the rest of this section values of the coupling constants
\begin{itemize}
\item $\Gs$, $\Go$, and $\Gr$ used for the RMF2 model are number $5$ of Table~\ref{tab:couplingsRMF2},
\item $\Gs$, $\Go$, $\Gr$, and $G_3$ used for the RMF3 model are number $15$ of Table~\ref{tab:couplingsRMF3}, and
\item $\Gs$, $\Go$, $\Gr$, $G_3$, and $G_4$ used for the RMF4 model are number $6$ of Table~\ref{tab:couplingsRMF4}.
\end{itemize}

\begin{figure}[H]
\centering
\scalebox{0.7}{\input{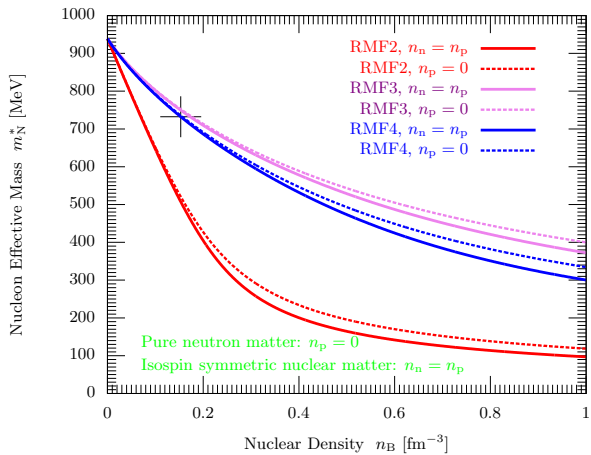}}
\caption{(Color online) The nucleon effective mass as a function of the nuclear density in the RMF approximation to the nonlinear ($\sigma-\omega$) models (RMF3 and  RMF4) and the linear model (RMF2) at zero temperature. The solid lines correspond to \ism, whereas dashed lines correspond to pure \nm.}
\label{fig:mnstarfig}
\end{figure}

Figure~\ref{fig:mnstarfig} shows the nucleon effective mass~\eqref{eq:mnstarInt} as a function of the nuclear density only for \ism\ and pure \nm. At low density, $\nb, \kf \to 0$, $\mn^*$ has almost the same behavior in the three models, because it converges to $\mn$ and then the nuclear medium has more tendency to an ideal gas behavior. Afterward, $\mn^*$ decreases almost linearly with the density. At higher density, above $0.6~\mathrm{fm^{-3}}$, it converges slower to zero in both RMF3 and RMF4 than in the RMF2 model; the first-order Taylor expansion of Eq.~\eqref{eq:mnstarInt} around $\mn^*/\mn$ yields
%
%
\begin{align} 
&\dfrac{\mn^{*}}{\mn}\left(n_{B}\gg n_{B}^{sat}\right)
\approx \\ \nonumber
&\dfrac{G_{3}G_{4}+G_{\sigma}\mn(G_{4}+G_{3}\mn)}{G_{3}G_{4}+G_{\sigma}\mn(2G_{4}+3G_{3}\mn)+G_{3}G_{4}G_{\sigma}\textcolor{red}{\kf^{2}}/\pi^{2}},
\end{align}
which gives in the RMF4 model $\mn^*\approx 238.3~\mathrm{MeV}$ at $\nb=1.0~\mathrm{fm^{-3}}$ and $\nn=\np$. Further, the behavior of $\mn^*$ for isospin symmetric nuclear and pure \nm\ is roughly the same at all densities owing to the approximative equality of the proton and neutron free masses and to the charge independence of the attractive force arising from the $\sigma$ meson exchange. At the saturation density~\eqref{eq:nbsat}, the coupling constants used here lead to $\mn^*\simeq 0.540\mn$ in RMF2, $0.798\mn$ in RMF3, and $0.780\mn$ in the RMF4 model.
%
%

The binding energy per nucleon is illustrated in Fig.~\ref{fig:energyfig} for both isospin symmetric nuclear and pure \nm; the difference between them (dot-dashed lines) is the asymmetry energy~\eqref{eq:asymeq2}. By way of comparison, we found that the curve for Eq.~\eqref{eq:asymeq} deviates by less than 2\% from that of Eq.~\eqref{eq:asymeq2}, which is numerically not significant. In all three models, values of $\nb^{sat}$, $E_{iso}^{sat}$, and $a^{sat}_{\text{sym}}$ are in perfect agreement with the empirical ones [Eqs.~\eqref{eq:nbsat}, \eqref{eq:energyV}, and \eqref{eq:asymV}], whereby the standard deviation errors are less than $10^{-4}\%$. All these deviations arise solely from the numerical solution of equation \eqref{eq:mnstarInt}. In RMF2, the saturation point $(\nb^{sat}, E_{iso}^{sat})$ is reproduced through $G_{\sigma}$ and $G_{\omega}$ only.
\begin{figure}[h]
\centering
\scalebox{0.7}{\input{energy_EoS.tex}}
\caption{(Color online) The binding energy per nucleon as a function of the nuclear density in the RMF approximation to the nonlinear ($\sigma-\omega$) models (RMF3 and  RMF4) and the linear model (RMF2) at zero temperature. The solid lines correspond to \ism, dashed lines to pure \nm, and dot-dashed lines to the asymmetry energy. The empirical values $\nb^{sat}$, $E_{iso}^{sat}$ and $a^{sat}_{\text{sym}}$ are reproduced with an standard deviation error less than $10^{-4}\%$.}
\label{fig:energyfig}
\end{figure}
%
%
\\For \ism\ at the saturation, bound states of nucleons appear with delicate cancellation between short-range repulsion and medium-range attraction. The system is in its ground state, whereby all energy levels are occupied separately by protons and neutrons up to the Fermi energy of about $779.3~\mathrm{MeV}$. Outside the saturation region, the system becomes unbound because the binding energy per nucleon becomes large (between $300$ and $500~\mathrm{MeV}$) comparing with the nucleon mass.

\begin{figure}[H]
\centering
\scalebox{0.7}{\input{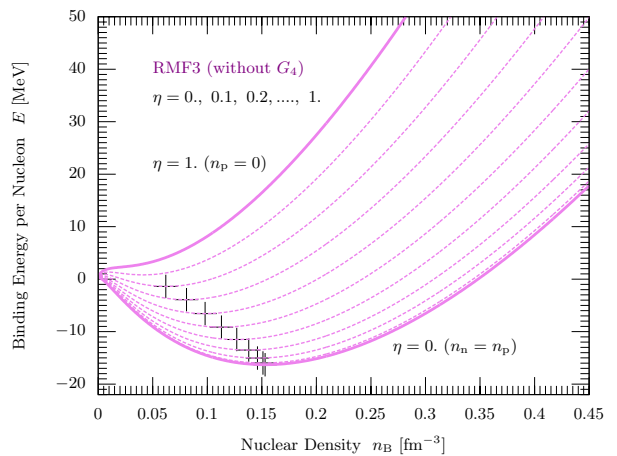}}
\caption{(Color online) The evolution of the saturation point with proton-neutron fraction $\eta$ in RMF3 model. The $\eta=0$ ($\eta=1$) solid curve corresponds to \ism\ (\nm). The $\eta=0.1,...,0.9$ dashed curves correspond to asymmetric nuclear matter which increases with $\eta$ from bottom to top. We found that nuclear matter ceases to saturate at $\eta$ slightly larger than $0.8$.}
\label{fig:energyasym}
\end{figure}
%
%
The evolution of the binding energy per nucleon of asymmetric nuclear matter with proton-neutron asymmetry $\eta$ is illustrated in Fig.~\ref{fig:energyasym} only for the nonlinear RMF3 model (for reasons that become clear in Sec.~\ref{sec:param}). The two solid lines correspond to isospin symmetric and pure \nm\ at $\eta=0$ and 1, respectively. The nine dashed lines correspond to asymmetric nuclear matter at $\eta=0.1,...,0.9$. We found that nuclear matter ceases to saturate at $\eta$ slightly larger than $0.8$, as shown in Table~\ref{tab:satpt}. By way of comparison, this value is slightly larger than that obtained by Piekarewicz and Centelles using the so-called NL3 and FSUGold model calculations~\cite{ref:PiekarewiczCentelles}. But the conclusion is the same, namely, the larger the proton-neutron asymmetry, the smaller the saturation density and hence the lower the binding energy per nucleon, which means that pure neutron matter is an unbound system at all densities.
\begin{table}[H]
\centering
\begin{tabular}{|l|l|l|l|}
\hline
\rule[0.6em]{0pt}{0.5em} & \multicolumn{3}{|c|}{Results for RMF3 (i.e., without $G_4$)} \\
\hline
\rule[0.6em]{0pt}{0.5em} P.-N. Asym. & \multicolumn{2}{|c|}{Saturation Point} & Compress. Mod.\\
\hline 
\rule[0.6em]{0pt}{0.5em} $\eta$ & $\nb^{sat}~[\mathrm{fm}^{-3}]$ & $E^{sat}~[\mathrm{MeV}]$ &  $\mathcal{K}~[\mathrm{MeV}]$ \\
\hline\hline
\rule[0.6em]{0pt}{0.5em} 0.0 ($\nn=\np$) & 0.153  & -16.3000 &  234.000~($=\mathcal{K}_{iso}$) \\
\hline
\rule[0.6em]{0pt}{0.5em} 0.1 & 0.151 & -15.9767 &  229.343 \\
\hline
\rule[0.6em]{0pt}{0.5em} 0.2 & 0.146 & -15.0260 &  220.058 \\
\hline
\rule[0.6em]{0pt}{0.5em} 0.3 & 0.138 & -13.5048 &  205.440 \\
\hline
\rule[0.6em]{0pt}{0.5em} 0.4 & 0.127 & -11.5035 &  184.452 \\
\hline
\rule[0.6em]{0pt}{0.5em} 0.5 & 0.113 & -9.14113 &  155.991 \\
\hline
\rule[0.6em]{0pt}{0.5em} 0.6 & 0.098 & -6.55846 &  128.252 \\
\hline
\rule[0.6em]{0pt}{0.5em} 0.7 & 0.081 & -3.91386 &  95.8663 \\
\hline
\rule[0.6em]{0pt}{0.5em} 0.8 & 0.062 & -1.38341 &  59.8410 \\
\hline
\rule[0.6em]{0pt}{0.5em} 0.9 & $\varnothing$ & $\varnothing$ &  (?) \\ 
\hline
\rule[0.6em]{0pt}{0.5em} 1.0 ($\np=0$) & $\varnothing$ & $\varnothing$ &  (?) \\
\hline
\end{tabular}
\caption{The evolution of the saturation density, the binding energy per nucleon and the compression modulus of nuclear matter with proton-neutron asymmetry in the nonlinear $\so$ model, without the quartic self-coupling constant $G_4$.}
\label{tab:satpt}
\end{table}

We close this section by discussing the behavior of the pressure and the compression modulus. For \ism, we obtain from Eq.~\eqref{eq:Ketaeq} $\mathcal{K}_{iso}\simeq556.693~\mathrm{MeV}$ in RMF2, which is much larger than the empirical one~\eqref{eq:kV}. This result is well known from the standard Walecka model~\cite{ref:Walecka}. In both RMF3 and RMF4 we obtain ``exactly what we want,'' $\mathcal{K}_{iso}\simeq234.0~\mathrm{MeV}$ owing (a) to the nonlinear extension known originally from Boguta and Bodmer~\cite{ref:BogutaBodmer}, (b) to the analytic expression for $\mathcal{K}$ given in Eq.~\eqref{eq:Ketaeq}, and (c) to the parametrization of the coupling constants, which would be presented in Sec.~\ref{sec:param}.\\

The compression modulus of asymmetric nuclear matter decreases with increasing proton-neutron asymmetry $\eta$ beween $0.0$ and $0.8$, as shown in the last column of Table~\ref{tab:satpt} for RMF3 model. Certain authors like Piekarewicz and Centelles~\cite{ref:PiekarewiczCentelles} obtained roughly the same results by expanding $\mathcal{K}(\eta)$ in a power series in $\eta^2$,
\begin{equation}
\nonumber
\mathcal{K}(\eta)=\mathcal{K}_{iso}+\eta^2\mathcal{K}_{\text{sym}}+\mathcal{O}(\eta^4),
\end{equation} 
which is fully analogous to the one for the binding energy per nucleon discussed earlier. Other authors like Blaizot \textit{et al.}~\cite{ref:Blaizot} write $\mathcal{K}(\eta)$ empirically as a sum of different types of compression modulus (volume, asymmetry, Coulomb, etc.), which can be done by fitting to the empirical data. In our case, the analytical expression~\eqref{eq:Ketaeq} includes (a) the compression modulus of \ism, which characterizes the small density fluctuations around the saturation point ($\nb^{sat},E^{sat}_{iso}$), and (b) the compression modulus of asymmetric matter, which characterizes the slope and curvature of the asymmetry energy around the saturation density at given value of $\eta$.

\begin{figure}[H]
\centering
\scalebox{0.7}{\input{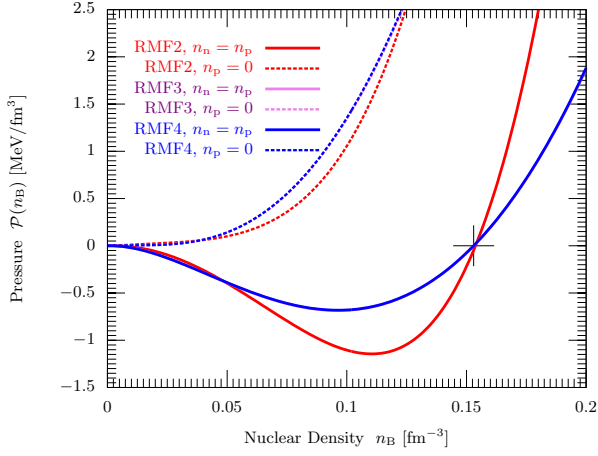}}
\caption{(Color online) The pressure as a function of the nuclear density in the RMF approximation to the nonlinear ($\sigma-\omega$) models (RMF3 and  RMF4) and the linear model (RMF2) at zero temperature. The solid lines correspond to \ism\ and dashed lines to pure \nm.}
\label{fig:pressurefig}
\end{figure}

In the case of a very asymmetric matter, $0.8<\eta\leqslant1$, the saturation point vanishes systematically, so that we cannot obtain a precise value for $\mathcal{K}$ within the $\so$ model (see Table~\ref{tab:satpt}). Nevertheless, we are ``curious'' to know the response of Eq.~\eqref{eq:Ketaeq} for some ``hypothetical equilibrium'' values of $\nb$ at $\eta=1$. This situation is quite similar to that of neutron stars in terms of the baryon density within an order of magnitude or less. The difference, however, lies in the fact that neutron stars are bound by gravity, not by the isospin symmetric nuclear forces. So, if we consider a superdense neutron matter at an ``equilibrium density'' between $0.8$ and $1.0~\mathrm{fm^{-3}}$, expression~\eqref{eq:Ketaeq} returns very high values of $\mathcal{K}=\mathcal{K}_{pnm}$ as presented in Table~\ref{tab:Kpnm}.
\begin{table}[H]
\centering
\begin{tabular}{|l|l|l|}
\hline
\rule[0.6em]{0pt}{0.5em} $\nb~[\mathrm{fm^{-3}}]$ & $\mn^*~[\mathrm{MeV}]$ & $\mathcal{K}_{pnm}~[\mathrm{MeV}]$ \\
\hline\hline
\rule[0.6em]{0pt}{0.5em} $0.8$ & $444.724$ & $8012.41$ \\
\hline
\rule[0.6em]{0pt}{0.5em} $0.9$ & $420.755$ & $9184.54$ \\
\hline
\rule[0.6em]{0pt}{0.5em} $1.0$ & $399.871$ & $10342.1$ \\
\hline
\end{tabular}
\caption{The compression modulus of pure \nm\ for RMF3 model for some ``hypothetical'' values of the equilibrium density.}
\label{tab:Kpnm}
\end{table}
Finally, Fig.~\ref{fig:pressurefig} shows the pressure for isospin symmetric nuclear (solid lines) and pure \nm\ (dashed lines). Its behavior in RMF3 as well as in RMF4 is almost the same owing to the value of the coupling constants used here.
%
%
At low density, the nuclear matter is in pure diluted phase, similar to a nonrelativistic ideal Fermi gas. In this case, expression~\eqref{eq:pressureInt} can be approximated by the following fifth-order polynomial of $\kf$
\begin{equation}
\label{eq:pressurelow}
\mathcal{P}^{\text{RMF4}}(n_B\ll \nb^{sat})\longrightarrow\dfrac{2}{15\pi^2}\dfrac{\kf^5}{\mn}.
\end{equation}
At higher density above the saturation, the pressure increases very strongly and the nuclear medium becomes in dense phase (liquid). In this case, expression~\eqref{eq:pressureInt} converges to
\begin{widetext}
\begin{eqnarray}
\label{eq:pressurehigh}
& &\hspace*{-2pc}
\mathcal{P}^{\text{RMF4}}(n_B\gg \nb^{sat})\longrightarrow\ \overbrace{\underbrace{\dfrac{1}{2}G_{\omega}\nb^2}_{\text{Pressure}\nearrow\nearrow\nearrow}\underbrace{\textcolor{red}{-\dfrac{\mn^2}{2G_{\sigma}}}\textcolor{violet}{-\dfrac{\mn^3}{3G_{3}}}\textcolor{blue}{-\dfrac{\mn^4}{4G_{4}}}}_{\text{Pressure}\searrow}\underbrace{\textcolor{red}{+\dfrac{\mn\mn^*}{G_{\sigma}}}\textcolor{violet}{+\dfrac{\mn^2\mn^*}{G_3}}\textcolor{blue}{+\dfrac{\mn^3\mn^*}{G_4}}}_{\text{correction 1st. order}\nearrow}}^{\text{Pressure}\nearrow\nearrow}+\underbrace{\mathcal{O}\left(\left(\dfrac{\mn^*}{\mn}\right)^{2}\right)}_{\text{negligible}}.
\end{eqnarray}
\end{widetext}
Here the leading term (which is the first) comes from the strong repulsive forces caused by the vector $\omega$ meson exchange. Afterward, small negative terms follow owing to attractive forces caused by the scalar $\sigma$ meson exchange (there is always exchange even at higher density), and the last three positive terms are their first relativistic corrections, which slightly increase the strength of the scalar interactions. Therefore, the second and third terms give the difference between the linear and the two nonlinear $\so$ models.
%
%
%
%
\section{Analytic Parametrization of the ($\sigma-\omega$) Model}
\label{sec:param}
First, let us collect the five empirical quantities for \ism\ discussed in the Introduction
\begin{subequations}
\label{eq:v}
\begin{align}
&\text{the saturation density:}~~\nb^{sat}=0.153~\mathrm{fm^{-3}},\label{eq:nb0}\\
&\text{the binding energy per nucleon:}~~ E_{iso}^{sat}=-16.3~\mathrm{MeV},\label{eq:en0}\\
&\text{the asymmetry energy coefficient:}~~ a^{sat}_{\text{sym}}=32.5 ~\mathrm{MeV}\label{eq:asym0},\\
&\text{the nucleon effective mass:}~~ 0.7 \lesssim \mn^*/\mn \lesssim 0.8,\label{eq:mnst0}\\
&\text{the compression modulus:}~~ 230.0 \lesssim \mathcal{K}_{iso} \lesssim 250.0~\mathrm{MeV}.\label{eq:Kiso0}
\end{align}
\end{subequations}
In this section, we answer the following questions
\begin{enumerate}[(i)]
\item Which values of the coupling constants of the $\so$ model do we have to input in the EoS in order to reproduce the five empirical quantities \eqref{eq:v}?
\item Is it possible to reproduce them without a four-body self-interaction?
\end{enumerate}
The input parameters for us here will be the five empirical quantities \eqref{eq:v}, while the output parameters will be
\begin{itemize}
\item $\Gs$, $\Go$, $\Gr$, $G_3$, and $G_4$ in RMF4,
\item $\Gs$, $\Go$, $\Gr$, and $G_3$ in RMF3, and
\item $\Gs$, $\Go$, and $\Gr$ in the RMF2 model.
\end{itemize}
Let us rewrite all EoS for cold \ism; after evaluating the integrals in Eqs.~\eqref{eq:mnstarInt}, \eqref{eq:pressureInt}, and \eqref{eq:energydInt}, we have the following.
\begin{widetext} 
\textit{The nucleon effective mass:}
\begin{eqnarray}
\label{eq:mnstar}
\mn^*=\mn-\frac{\Gs}{\pi^2}\mn^*\left[\ef^*\kf-\mn^{*2} \ln{\left(\frac{\ef^*+\kf}{\mn^*}\right)}\right]+\frac{\Gs}{G_3}(\mn-\mn^*)^2+\frac{\Gs}{G_4}(\mn-\mn^*)^3
\end{eqnarray}
\textit{The pressure}
\begin{subequations}
\label{eq:pressured}
\begin{align}
\label{eq:pressuredRMF2}
&\textcolor{red}{\mathcal{P}^{RMF2}=\dfrac{1}{2}G_{\omega}\nb^2-\frac{(\mn-\mn^*)^2}{2 \Gs}+} \textcolor{red}{\dfrac{1}{4\pi^2}\left[\dfrac{2}{3}\ef^*\kf^{*3}-\ef^*\kf\mn^{*2}+\mn^{*4}\ln\left(\dfrac{E_{F}^{*}+\textbf{\textit{p}}_{F}}{\mn^*}\right)\right],}\\
\label{eq:pressuredRMF3}
&\textcolor{violet}{\mathcal{P}^{RMF3}=}~\textcolor{red}{\mathcal{P}^{RMF2}}\textcolor{violet}{-\dfrac{(\mn-\mn^*)^3}{3G_{3}},}\\
\label{eq:pressuredRMF4}
&\textcolor{blue}{\mathcal{P}^{RMF4}=}~\textcolor{violet}{{\mathcal{P}^{RMF3}}}\textcolor{blue}{-\dfrac{(\mn-\mn^*)^4}{4G_{4}}.}
\end{align}
\end{subequations}
\textit{The energy density}
\begin{subequations}
\label{eq:energyd}
\begin{align}
\label{eq:energydRMF2}
&\textcolor{red}{\mathcal{E}^{RMF2}=\dfrac{1}{2}G_{\omega}\nb^2+\frac{(\mn-\mn^*)^2}{2 \Gs}+} \textcolor{red}{\dfrac{1}{4\pi^2}\left[2\ef^{*3}\kf-\ef^*\kf\mn^{*2}-\mn^{*4}\ln\left(\dfrac{E_{F}^{*}+\textbf{\textit{p}}_{F}}{\mn^*}\right)\right],}\\
\label{eq:energydRMF3}
&\textcolor{violet}{\mathcal{E}^{RMF3}=}~\textcolor{red}{\mathcal{E}^{RMF2}}\textcolor{violet}{+\dfrac{(\mn-\mn^*)^3}{3G_{3}},}\\
\label{eq:energydRMF4}
&\textcolor{blue}{\mathcal{E}^{RMF4}=}~\textcolor{violet}{{\mathcal{E}^{RMF3}}}\textcolor{blue}{+\dfrac{(\mn-\mn^*)^4}{4G_{4}}.}
\end{align}
\end{subequations}
\textit{The asymmetry energy}
\begin{equation}
\label{eq:asym}
a_{\text{sym}}=\dfrac{\kf^2}{6\ef^*}+\dfrac{1}{8}\Gr\nb.
\end{equation} 
\textit{The compression modulus}
\begin{eqnarray}
\label{eq:Kiso}
\mathcal{K}_{iso}=\frac{3 \kf^2}{\ef^*}+\frac{6\Go}{\pi^2}\kf^3-\frac{6\Gs\kf^3\mn^{*2}}{\pi^2\ef^{*2}
}\left[1+\Gs\left[\frac{2 \sigma_0}{G_3}+\frac{3 \sigma_0^2}{G_4}+\frac{2}{\pi
^2}\left(\frac{\kf^3+3\kf\mn^{*2}}{2 \ef^*}-\frac{3}{2} \mn^{*2}\ln{\left(\frac{\ef^*+\kf}{\mn^*}\right)}\right)\right]\right]^{-1},
\end{eqnarray}
where $\mathcal{K}_{iso}$ is deduced from~\eqref{eq:Ketaeq} for $\eta=0$. Further we have $$\nb=\dfrac{2}{3\pi^2}\kf^3,~~~~~E_{iso}\equiv \dfrac{B_{iso}}{A}=\dfrac{\mathcal{E}_{iso}^{RMF4}}{\nb}-\mn,~~~~~\sigma_0=\mn-\mn^*~~~~~\text{and}~~~\ef^*\equiv\mu^*_0=\sqrt{\kf^2+\mn^{*2}}.$$
\end{widetext} 
In the next three sections, all quantities $\nb$, $E_{iso}$, $\kf$, $\mn^*$, $\ef^*$, and $a_{\text{sym}}$ are evaluated at the saturation ($\nb=\nb^{sat}, E_{iso}=E_{iso}^{sat}$, etc.).
%
%
\subsection{Parametrization of the RMF4 Model}
\label{subsec:rmf4}
The isospin coupling constant can be immediately obtained from Eq.~\eqref{eq:asym} as
\begin{eqnarray}
\label{eq:Gr}
\Gr=\frac{8}{\nb}\left(a_{\text{sym}}-\frac{\kf^2}{6 \ef^*}\right).
\end{eqnarray}
By adding the EoS \eqref{eq:pressuredRMF4} to Eq.~\eqref{eq:energydRMF4} and requiring that $\mathcal{P}\stackrel{!}{=}0~\mathrm{MeV/fm^3}$, we obtain the repulsive coupling constant
\begin{eqnarray}
\label{eq:Go}
\Go=\frac{E_{iso}+\mn-\ef^*}{\nb}.
\end{eqnarray}
Now, to determine $\Gs$, $G_3$, and $G_4$, we need three equations, for example, Eqs.~\eqref{eq:mnstar}, \eqref{eq:Kiso}, and \eqref{eq:energydRMF4} [or \eqref{eq:pressuredRMF4}]. From Eqs.~\eqref{eq:mnstar} and \eqref{eq:energydRMF4}, we obtain
\begin{align}
\label{eq:Gse1}
\dfrac{ce3}{G_3}+\dfrac{ce4}{G_4}=ce5,
\end{align}
with
\begin{align}
\nonumber
&ce3=\dfrac{\sigma_0^3}{6}, \\ \nonumber
&ce4=\dfrac{\sigma_0^4}{4}, \\ \nonumber
&ce5=\dfrac{1}{2}\sigma_0\ns+ce2-\mathcal{E}^{RMF4}, \\ \nonumber
&ce2=\dfrac{1}{2}\Go\nb^2+\\ \nonumber
&\dfrac{1}{4\pi^2}\left[2\ef^{*3}\kf-\ef^*\kf\mn^{*2}-\mn^{*4}\ln\left(\dfrac{E_{F}^{*}+\textbf{\textit{p}}_{F}}{\mn^*}\right)\right], \\ \nonumber
&\ns=\frac{\mn^*}{\pi^2}\left[\ef^*\kf-\mn^{*2} \ln{\left(\frac{\ef^*+\kf}{\mn^*}\right)}\right].
\end{align}
Further, from Eqs.~\eqref{eq:mnstar}, and \eqref{eq:Kiso} we obtain
\begin{align}
\label{eq:Gse2}
\dfrac{\sigma_0}{G_3}+\dfrac{2\sigma_0^2}{G_4}=c16,
\end{align}
with
\begin{align}
\nonumber
&c16=\dfrac{c14}{c15}-\dfrac{2}{\pi^2}c13-\dfrac{\ns}{\sigma_0}, \\ \nonumber
&c15=\dfrac{6}{\pi^2}\Go\kf^3+\dfrac{3\kf^2}{\ef^*}-\mathcal{K}_{iso}, \\ \nonumber
&c14=\dfrac{6\kf^3}{\pi^2}\left(\dfrac{\mn^*}{\ef^*}\right)^2, \\ \nonumber
&c13=\dfrac{\kf^3+3\kf\mn^{*2}}{2\ef^*}-\dfrac{3}{2}\mn^{*2}\cdot\ln\left(\dfrac{\kf+\ef^*}{\mn^*}\right).
\end{align}
Finally, from Eqs.~\eqref{eq:Gse1} and \eqref{eq:Gse2} we deduce $\Gs$, $G_3$ and $G_4$ in terms of the empirical quantities $\nb$, $E_{iso}$, $\mn^*$, and $\mathcal{K}_{iso}$
%
%
\begin{widetext}
\begin{eqnarray}
\label{eq:Gs}
& & \hspace*{-1pc}
\Gs=\Bigg\{\pi^2\ef^*(\mn-\mn^*)^2 \left[\pi^2\left(3 \kf^2-\ef^*\mathcal{K}_{iso}\right)-\frac{6}{\nb}\ef^*\kf^3\big(\ef^*-E_{iso}-\mn\big) \right]\Bigg\}
\Bigg /
\nonumber \\ & & \hspace*{1pc}
\Bigg\{ 6 \ef^{*2} \kf (\mn-\mn^*) \mn^* \left[\frac{6}{\nb}\ef^*\kf^3\big(\ef^*-E_{iso}-\mn\big)-\pi^2\left(3\kf^2-\ef^*\mathcal{K}_{iso}\right)\right]+
\nonumber \\ & & \hspace*{1pc}
\ef^*\left[\pi^2\left(3\kf^2-\ef^*\mathcal{K}_{iso}\right)-\frac{6}{\nb}\ef^*\kf^3\big(\ef^*-E_{iso}-\mn\big)\right]\bigg[6\pi^2\nb(E_{iso}+\mn)+\ef^*\big(2 \kf^3-3\kf
\mn^{*2}-6\pi^2\nb\big)\bigg]+
\nonumber \\ & & \hspace*{1pc}
\kf(\mn-\mn^*)^2 \left[\frac{\ef^*}{\nb}\left(\kf^2+3 \mn^{*2}\right)\bigg(6\ef^*\kf^3+\pi^2\nb\mathcal{K}_{iso}-6 \kf^3 (E_{iso}+\mn)
\bigg)-3\left(\pi\kf\ef^*\right)^2\right]+
\nonumber \\ & & \hspace*{1pc}
3\ef^* \mn^2 \mn^{*2} \left[\pi^2\big(3\kf^2-\ef^* \mathcal{K}_{iso}\big)-\frac{6}{\nb}\ef^*\kf^3\big(\ef^*-E_{iso}-\mn\big)\right] \ln{\left(\frac{\ef^*+\kf}{\mn^*}\right)}\Bigg\}.
\end{eqnarray}
%
%
\begin{eqnarray}
\label{eq:G3}
& & \hspace*{-1pc}
G_3=\Bigg\{\pi^2\ef^*(\mn-\mn^*)^3\left[\pi^2\big(3\kf^2-\ef^*\mathcal{K}_{iso}\big)-\frac{6}{\nb}\ef^*\kf^3\big(\ef^*-E_{iso}-\mn\big)\right]\Bigg\}\Bigg/
\nonumber \\ & & \hspace*{1pc}
\Bigg\{15 \ef^{*2} \kf (\mn-\mn^*) \mn^*\left[\pi^2\left(3 \kf^2-\ef^*\mathcal{K}_{iso}\right)-\frac{6}{\nb}\ef^*\kf^3\big(\ef^*-E_{iso}-\mn\big) \right]+
\nonumber \\ & & \hspace*{1pc}
3\kf(\mn-\mn^*)^2\Bigg[3\left(\pi\kf\ef\right)^2
+\ef^* \left(\kf^2+3 \mn^{*2}\right)\left(\frac{6}{\nb}\kf^3(E_{iso}+\mn-\ef^*)-\pi^2\mathcal{K}_{iso}
\right)\Bigg]+
\nonumber \\ & & \hspace*{1pc}
2 \ef^*\Bigg[\frac{6}{\nb}\ef^*\kf^3\big(\ef^*-E_{iso}-\mn\big)-\pi^2\left(3\kf^2-\ef^*\mathcal{K}_{iso}\right)\Bigg] \Bigg[6\pi^2\nb(E_{iso}+\mn) +\ef^*\bigg(2\kf^3-3 \kf \mn^{*2}-6\pi^2\nb\bigg)\Bigg]+
\nonumber \\ & & \hspace*{1pc}
3\mn\ef^*\mn^{*2}(3 \mn-\mn^*)\bigg[\dfrac{6}{\nb}\ef^{*2}\kf^3-\dfrac{6}{\nb}\ef^*
\kf^3 (E_{iso}+\mn)-\pi^2\left(3\kf^2-\ef^*\mathcal{K}_{iso}\right)\bigg]\ln{\left(\frac{\ef^*+\kf}{\mn^*}\right)}\Bigg\}.
\end{eqnarray}
%
%
\begin{eqnarray}
\label{eq:G4}
& & \hspace*{-1.3pc}
G_4=\Bigg\{\pi^2\ef^*(\mn-\mn^*)^4\left[\pi ^2\left(3\kf^2-\ef^*\mathcal{K}_{iso}\right)-\frac{6}{\nb}\ef^*\kf^3\big(\ef^*-E_{iso}-\mn\big)\right]\Bigg\}
\Bigg/
\nonumber \\ & & \hspace*{0.75pc}
\Bigg\{8\ef^{*2}\kf(\mn-\mn^*)\mn^*\bigg[\frac{6}{\nb}\ef^*\kf^3 \big(\ef^*-E_{iso}-\mn\big)-3\pi^2\kf^2 +\pi^2\ef^*\mathcal{K}_{iso} \bigg]-
\nonumber \\ & & \hspace*{1pc}
2\kf(\mn-\mn^*)^2\bigg[3\left(\pi\kf\ef\right)^2+\ef^* \left(\kf^2+3 \mn^{*2}\right)\left(\frac{6}{\nb}\kf^3 (E_{iso}-\ef^*+\mn)-\pi ^2\mathcal{K}_{iso}\right)\bigg]+
\nonumber \\ & & \hspace*{1pc}
\ef^*\bigg[\pi^2\left(3 \kf^2-\ef^*\mathcal{K}_{iso}\right)-\frac{6}{\nb}\ef^*\kf^3\big(\ef^*-E_{iso}-\mn\big)\bigg]\bigg[6\pi ^2\nb(E_{iso}+\mn)+\ef^*\bigg(2\kf^3-3\kf\mn^{*2}-6\pi^2\nb\bigg)\bigg]-
\nonumber \\ & & \hspace*{0.6pc}
\ef^*\mn^{*2}\bigg(6 \mn^2-4 \mn \mn^*+\mn^{*2}\bigg)\bigg[\dfrac{6}{\nb}\ef^{*2}
\kf^3-\dfrac{6}{\nb}\ef^*\kf^3 (E_{iso}+\mn)-3\pi^2\kf^2+\pi^2\ef^*\mathcal{K}_{iso}\bigg]\ln{\left(\frac{\ef^*+\kf}{\mn^*}\right)}\Bigg\}.
\end{eqnarray}
\end{widetext}
As we see, they are expressed as rational functions, $\{...\}/\{...\}$. The relation between $\Gs$ and the nonlinear coupling constants is given by
\begin{equation}
\label{eq:Gs34}
\Gs=\dfrac{\sigma_0}{\ns-\sigma_0^2/G_3-\sigma_0^3/G_4}.
\end{equation} 
The results are shown in Table~\ref{tab:couplingsRMF4} for different values of $\mathcal{K}_{iso}$ between $230$ and $250~\mathrm{MeV}$ and $\mn^*/\mn$ between $0.7$ and $0.8$. In each line, the set of coupling constants yields the saturation density \eqref{eq:nb0}, the binding energy per nucleon at saturation \eqref{eq:en0} and the asymmetry energy coefficient \eqref{eq:asym0} with standard deviation errors less than $10^{-4}\%$ arising from the numerical solution of equation Eq.~\eqref{eq:mnstar} only. Thus, $G_3$ and $G_4$ together provide high accuracy in determining of the five empirical quantities~\eqref{eq:v}. This accuracy was also confirmed by other authors~\cite{ref:Lalazissis, ref:ToddRutel} by investigating the ground state properties of finite nuclei through accurately calibrating procedures, such as the NL3 and FSUGold models.
\\However, for $230.0\leqslant \mathcal{K}_{iso} \leqslant 232.7~\mathrm{MeV}$, we found that the quartic self-coupling constant $G_4$ is negative at all values of $\mn^*/\mn$. Further, at $\mathcal{K}_{iso}=234.0~\mathrm{MeV}$ (the value that we have supported in this work \cite{ref:MyersSwiatecki1998}) and for $\mn^*/\mn\lesssim 0.79$, $G_4$ is always negative, while for $\mn^*/\mn>0.79$ it takes very large positive values (above $1987.3$). Notice that $G_4$ should be less than $1000$ to have an effect of roughly $22.1\%$ on the binding energy per nucleon at the saturation. Here we found that it is only possible to get $G_4$ between $0$ and $1000$ if $0.731\lesssim \mn^*/\mn \lesssim 0.8$ and $235.3\lesssim \mathcal{K}_{iso} \lesssim 250.0~\mathrm{MeV}$. Of course, mathematically it does not matter if $G_4$ is negative or positive, because all sets of coupling constants in Table~\ref{tab:couplingsRMF4}, which fulfill Eqs.~\eqref{eq:Gr}, \eqref{eq:Gse2} and \eqref{eq:Gs34} at saturation, yield the same behavior of the EoS at all nuclear densities (see Fig.s~\ref{fig:mnstarfig},~\ref{fig:energyfig}, and~\ref{fig:pressurefig} for RMF4, where $G_4=-254.991$).
%
%
%
%
\begin{table}[H]
\centering
%
\begin{tabular}{|l|l|l|l|l|l|l|l|}
 \hline
\rule[0.6em]{0pt}{0.5em} \multirow{2}{*}{Nr.} & \multirow{2}{*}{$\dfrac{\mn^*}{\mn}$} & $\mathcal{K}_{iso}$ & $G_{\sigma}$ & $G_{\omega}$ & $G_{\rho}$ & $G_{3}$ & $G_{4}$ \\
\rule[0.6em]{0pt}{0.5em} &  & \tiny{$[\mathrm{MeV}]$} & \tiny{$[\mathrm{GeV}^{-2}]$} & \tiny{$[\mathrm{GeV}^{-2}]$} & \tiny{$[\mathrm{GeV}^{-2}]$} & \tiny{$[\mathrm{GeV}^{-1}]$}& \\
\hline\hline
\rule[0.6em]{0pt}{0.5em}    1  &   0.71  &   230.0  &   311.430  &   176.413  &   114.672  &   204.403  &  -182.178   \\
\hline
\rule[0.6em]{0pt}{0.5em}    2  &   0.71  &   232.0  &   310.946  &   176.413  &   114.672  &   206.731  &  -186.768   \\
\hline
\rule[0.6em]{0pt}{0.5em}    3  &   0.72  &   230.0  &   303.947  &   168.962  &   115.960  &   188.641  &  -174.967   \\
\hline
\rule[0.6em]{0pt}{0.5em}    4  &   0.72  &   232.0  &   303.442  &   168.962  &   115.960  &   190.892  &  -179.960   \\
\hline
\rule[0.6em]{0pt}{0.5em}    5  &   0.72  &   234.0  &   302.937  &   168.962  &   115.960  &   193.204  &  -185.258   \\
\hline
\rule[0.6em]{0pt}{0.5em}    \textcolor{blue}{6}  &   \textcolor{blue}{0.78}  &   \textcolor{blue}{234.0}  &   \textcolor{blue}{257.169}  &   \textcolor{blue}{123.993}  &   \textcolor{blue}{123.119}  &   \textcolor{blue}{116.736}  &  \textcolor{blue}{-254.991}   \\
\hline
\rule[0.6em]{0pt}{0.5em}    7  &   0.73  &   230.0  &   296.465  &   161.498  &   117.220  &   173.707  &  -169.047   \\
\hline
\rule[0.6em]{0pt}{0.5em}    8  &   0.73  &   232.0  &   295.936  &   161.498  &   117.220  &   175.887  &  -174.583   \\
\hline
\rule[0.6em]{0pt}{0.5em}    9  &   0.73  &   234.0  &   295.408  &   161.498  &   117.220  &   178.126  &  -180.508   \\
\hline
\rule[0.6em]{0pt}{0.5em}    10  &   0.73  &   236.0  &   294.880  &   161.498  &   117.220  &   180.429  &  -186.863   \\
\hline
\rule[0.6em]{0pt}{0.5em}   11  &   0.74  &   230.0  &   288.972  &   154.021  &   118.452  &   159.642  &  -164.922   \\
\hline
\rule[0.6em]{0pt}{0.5em}   12  &   0.74  &   232.0  &   288.417  &   154.021  &   118.452  &   161.755  &  -171.233   \\
\hline
\rule[0.6em]{0pt}{0.5em}   13  &   0.74  &   234.0  &   287.863  &   154.021  &   118.452  &   163.930  &  -178.063   \\
\hline
\rule[0.6em]{0pt}{0.5em}   14  &   0.74  &   236.0  &   287.310  &   154.021  &   118.452  &   166.170  &  -185.479   \\
\hline
\rule[0.6em]{0pt}{0.5em}   15  &   0.75  &   230.0  &   281.456  &   146.532  &   119.657  &   146.478  &  -163.434   \\
\hline
\rule[0.6em]{0pt}{0.5em}   16  &   0.75  &   232.0  &   280.872  &   146.532  &   119.657  &   148.533  &  -170.930   \\
\hline
\rule[0.6em]{0pt}{0.5em}   17  &   0.75  &   234.0  &   280.289  &   146.532  &   119.657  &   150.653  &  -179.168   \\
\hline
%
\rule[0.6em]{0pt}{0.5em}   18  &   0.75  &   236.0  &   279.706  &   146.532  &   119.657  &   152.839  &  -188.264   \\
\hline
\rule[0.6em]{0pt}{0.5em}   19  &   0.76  &   230.0  &   273.900  &   139.030  &   120.836  &   134.247  &  -166.160   \\
\hline
\rule[0.6em]{0pt}{0.5em}   20  &   0.76  &   232.0  &   273.283  &   139.030  &   120.836  &   136.256  &  -175.643   \\
\hline
\rule[0.6em]{0pt}{0.5em}   21  &   0.76  &   234.0  &   272.666  &   139.030  &   120.836  &   138.332  &  -186.304   \\
\hline
\rule[0.6em]{0pt}{0.5em}   22  &   0.77  &   230.0  &   266.280  &   131.517  &   121.990  &   122.983  &  -176.534   \\
\hline
\rule[0.6em]{0pt}{0.5em}   23  &   0.77  &   232.0  &   265.625  &   131.517  &   121.990  &   124.961  &  -189.847   \\
\hline
\rule[0.6em]{0pt}{0.5em}   24  &   0.79  &   246.0  &   244.787  &   116.459  &   124.225  &   122.098  &   682.043   \\
\hline
\rule[0.6em]{0pt}{0.5em}   25  &   0.79  &   248.0  &   244.056  &   116.459  &   124.225  &   124.937  &   477.204   \\
\hline
\rule[0.6em]{0pt}{0.5em}   26  &   0.79  &   250.0  &   243.327  &   116.459  &   124.225  &   127.922  &   366.713   \\
\hline
\rule[0.6em]{0pt}{0.5em}   27  &   0.80  &   236.0  &   240.243  &   108.914  &   125.307  &   101.935  &   777.923   \\
\hline
\rule[0.6em]{0pt}{0.5em}   28  &   0.80  &   238.0  &   239.447  &   108.914  &   125.307  &   104.286  &   482.989   \\
\hline
\rule[0.6em]{0pt}{0.5em}   29  &   0.80  &   240.0  &   238.654  &   108.914  &   125.307  &   106.756  &   349.882   \\
\hline
\rule[0.6em]{0pt}{0.5em}   30  &   0.80  &   242.0  &   237.863  &   108.914  &   125.307  &   109.355  &   274.087   \\
\hline
\rule[0.6em]{0pt}{0.5em}   31  &   0.80  &   244.0  &   237.074  &   108.914  &   125.307  &   112.093  &   225.146   \\
\hline
\rule[0.6em]{0pt}{0.5em}   32  &   0.80  &   246.0  &   236.288  &   108.914  &   125.307  &   114.982  &   190.935   \\
\hline
\rule[0.6em]{0pt}{0.5em}   33  &   0.80  &   248.0  &   235.505  &   108.914  &   125.307  &   118.035  &   165.674   \\
\hline
\rule[0.6em]{0pt}{0.5em}   34  &   0.80  &   250.0  &   234.724  &   108.914  &   125.307  &   121.266  &   146.257   \\
\hline
\end{tabular}
\caption{Coupling constants of the nonlinear $\so$ model for nuclear matter, RMF4, for different values of the nucleon effective mass and the compression modulus, that reproduce the saturation density, the binding energy per nucleon at saturation and the asymmetry energy coefficient.}
\label{tab:couplingsRMF4}
\end{table}
%
Although, negative values of $G_4$ do not constitute good news for the nonlinear $\so$ model, we should not dramatize this for the reason that the scalar $\sigma$ field is treated as a classical mean field, which is not realistic. Further, the $\so$ model treats baryons and mesons as elementary degrees of freedom; this can be one of the reasons why various values of $G_4$ are negative at given values of $\mathcal{K}_{iso}$ and $\mn^*$. To be closer to realistic descriptions, the coupling constants should, for example, depend on the momentum transfer between nucleons. In other words, negative coupling constants will be a serious problem when (a) we examine the $\so$ model with momentum-dependent coupling constants of the meson fields, that is, by incorporating form factors into the nucleon-meson couplings, or (b) we derive the EoS for nuclear matter directly from QCD, which is, however, not achieved so far.
%
%
\subsection{Parametrization of the RMF3 Model}
\label{subsec:rmf3}
In RMF3 model we take $G_4\longrightarrow \pm \infty$. To give a positive answer to the second question of this section, the coupling constant $G_3$ must
\begin{itemize}
\item verify simultaneously Eqs.~\eqref{eq:Gse1} and~\eqref{eq:Gse2}, which is the best condition, or
\item verify Eq.~\eqref{eq:Gse1} and ``approximately'' Eq.~\eqref{eq:Gse2}, or
\item verify Eq.~\eqref{eq:Gse2} and ``approximately'' Eq.~\eqref{eq:Gse1}.
\end{itemize}
If $G_3\stackrel{!}{=}ce3/ce5$, then $\mn^*$ and $E_{iso}$ will be reproduced, but there is no guarantee for $\mathcal{K}_{iso}$. If $G_3\stackrel{!}{=}\sigma_0/c16$, then $\mn^*$ and $\mathcal{K}_{iso}$ will be reproduced, but there is no guarantee for $E_{iso}$. Thus, let us define the numerical parameter
\begin{equation}
\label{eq:diffe}
\verb|diffe|\stackrel{\text{def}}{=}\Big|\dfrac{\sigma_0}{c16}-\dfrac{ce3}{ce5}\Big|.
\end{equation} 
So, if $\verb|diffe|\longrightarrow 0$ and $G_3$ verify Eq.~\eqref{eq:Gse1} or~\eqref{eq:Gse2}, then the five empirical quantities~\eqref{eq:v} can well be reproduced in RMF3 model. The smaller the parameter $\verb|diffe|$ is, the more precisely we obtain our desired results. In Table~\ref{tab:couplingsRMF3}, we present some results for $\verb|diffe|\leqslant 10^{-7}$ and by requiring that $G_3\stackrel{!}{=}ce3/ce5$.
%
%
%
%
%
%
%
%
\begin{table}[H]
\centering
%
\begin{tabular}{|l|l|l|l|l|l|l|l|}
\hline
\rule[0.6em]{0pt}{0.5em} \multirow{2}{*}{Nr.} & \multirow{2}{*}{$\dfrac{\mn^*}{\mn}$} & $\mathcal{K}_{iso}$ & $G_{\sigma}$ & $G_{\omega}$ & $G_{\rho}$ & $G_{3}$ \\
\rule[0.6em]{0pt}{0.5em} &  & $[\mathrm{MeV}]$ & $[\mathrm{GeV}^{-2}]$ & $[\mathrm{GeV}^{-2}]$ & $[\mathrm{GeV}^{-2}]$ & $[\mathrm{GeV}^{-1}]$ \\
\hline\hline
%
\rule[0.6em]{0pt}{0.5em}    1  &     0.775  &   253.862  &   254.463  &   127.757  &   122.558  &   147.074    \\
 \hline
\rule[0.6em]{0pt}{0.5em}    2  &     0.780  &   249.712  &   251.748  &   123.993  &   123.119  &   136.037    \\
 \hline
\rule[0.6em]{0pt}{0.5em}    3  &     0.781  &   248.878  &   251.212  &   123.240  &   123.231  &   133.913    \\
 \hline
\rule[0.6em]{0pt}{0.5em}    4  &     0.782  &   248.043  &   250.679  &   122.487  &   123.342  &   131.815    \\
\hline
\rule[0.6em]{0pt}{0.5em}    5  &     0.784  &   246.368  &   249.619  &   120.981  &   123.564  &   127.699    \\
 \hline
\rule[0.6em]{0pt}{0.5em}    6  &     0.786  &   244.687  &   248.569  &   119.474  &   123.785  &   123.687    \\
 \hline
\rule[0.6em]{0pt}{0.5em}    7  &     0.787  &   243.844  &   248.048  &   118.720  &   123.896  &   121.718    \\
 \hline
\rule[0.6em]{0pt}{0.5em}    8  &     0.788  &   242.999  &   247.530  &   117.967  &   124.005  &   119.775    \\
 \hline
\rule[0.6em]{0pt}{0.5em}    9  &     0.790  &   241.304  &   246.502  &   116.459  &   124.225  &   115.963    \\
 \hline
\rule[0.6em]{0pt}{0.5em}    10  &     0.793  &   238.746  &   244.983  &   114.196  &   124.552  &   110.426    \\
 \hline
\rule[0.6em]{0pt}{0.5em}    11  &     0.794  &   237.889  &   244.483  &   113.442  &   124.660  &   108.627    \\
 \hline
\rule[0.6em]{0pt}{0.5em}    12  &     0.795  &   237.030  &   243.986  &   112.687  &   124.769  &   106.851    \\
 \hline
\rule[0.6em]{0pt}{0.5em}    13  &     0.797  &   235.305  &   243.003  &   111.178  &   124.985  &   103.368    \\
 \hline
\rule[0.6em]{0pt}{0.5em}    14  &     0.798  &   234.450  &   242.521  &   110.424  &   125.092  &   101.656    \\
 \hline
\rule[0.6em]{0pt}{0.5em}    \textcolor{violet}{15}  &     \textcolor{violet}{0.798}   &  \textcolor{violet}{234.001}   &  \textcolor{violet}{242.276}  &  \textcolor{violet}{110.046}   &   \textcolor{violet}{125.146}  &   \textcolor{violet}{100.815}    \\
 \hline
\rule[0.6em]{0pt}{0.5em}    16  &     0.799  &   233.570  &   242.033  &   109.669  &   125.200  &   99.975    \\
 \hline
\rule[0.6em]{0pt}{0.5em}    17  &     0.800  &   232.710  &   241.559  &   108.914  &   125.307  &    98.307   \\
\hline
\end{tabular}
\caption{Coupling constants of the nonlinear $\so$ model, RMF3, for different values of the nucleon effective mass and the compression modulus, that reproduce the saturation density, the binding energy per nucleon at saturation and the asymmetry energy coefficient.}
\label{tab:couplingsRMF3}
\end{table}
%
Certainly we can obtain roughly the same sets of coupling constants by using Eqs.~\eqref{eq:mnstar}, \eqref{eq:Kiso}, and \eqref{eq:pressuredRMF4} [instead of Eq.~\eqref{eq:energydRMF4}], and then we can parametrize the RMF3 model by defining the following numerical parameter:
\begin{equation}
\label{eq:diffp}
\verb|diffp|\stackrel{\text{def}}{=}\Big|\dfrac{\sigma_0}{c16}-\dfrac{cp3}{cp5}\Big|,
\end{equation}
where
\begin{align}
\nonumber
&cp5=\dfrac{2}{\sigma_0^2}cp2-\dfrac{\ns}{\sigma_0}, \\ \nonumber
&cp3=-\dfrac{\sigma_0}{3}, \\ \nonumber
&cp2=\dfrac{1}{2}\Go\nb^2+\\ \nonumber
&\dfrac{1}{4\pi^2}\left[\dfrac{2}{3}\ef^*\kf^3-\ef^*\kf\mn^{*2}-\mn^{*4}\ln\left(\dfrac{E_{F}^{*}+\textbf{\textit{p}}_{F}}{\mn^*}\right)\right].
\end{align}
Finally, if we compare the results of Table~\ref{tab:couplingsRMF4} with those of Table~\ref{tab:couplingsRMF3}, we see that the accuracy in terms of the nucleon effective mass and/or the compression modulus obtained earlier in RMF4 becomes more difficult in the RMF3 model. This is the price we have to pay in RMF3, which is of minor importance as long as the empirical values of $\mn^*$ and $\mathcal{K}$ cannot be estimated (at present) to an accuracy of better than $\sim 0.02\mn$ and $\sim 0.2~\mathrm{MeV}$, respectively.
%
%
%
\subsection{Parametrization of the RMF2 Model}
\label{subsec:rmf2}
To be in the linear $\so$ model, we can take $G_3\longrightarrow +\infty$ and $G_4\longrightarrow \pm\infty$. The scalar coupling constant becomes $\Gs=\sigma_0/\ns$. This, together with $\Go$ and $\Gr$, yields just $\nb$, $E_{iso}$ and $a_{\text{sym}}$ at saturation. Some values are listed in Table~\ref{tab:couplingsRMF2}.
%
%
%
\begin{table}[H]
\centering
\begin{tabular}{|l|l|l|l|l|l|l|l|}
 \hline
\rule[0.6em]{0pt}{0.5em} Nr. & $\mn^*/\mn$ &  $\mathcal{K}_{iso}[\mathrm{MeV}]$& $G_{\sigma}[\mathrm{GeV}^{-2}]$ & $G_{\omega}[\mathrm{GeV}^{-2}]$ & $G_{\rho}[\mathrm{GeV}^{-2}]$ \\
\hline\hline
\rule[0.6em]{0pt}{0.5em}     1  &  0.536  & 576.368 &   398.114  &   303.302  &    86.608    \\
 \hline
\rule[0.6em]{0pt}{0.5em}     2  &  0.537  & 570.700 &   397.293  &   302.592  &    86.806    \\
 \hline
\rule[0.6em]{0pt}{0.5em}     3  &  0.538  & 565.052 &   396.472  &   301.881  &    87.004    \\
 \hline
\rule[0.6em]{0pt}{0.5em}     4  &  0.539  & 559.441 &   395.651  &   301.171  &    87.201    \\
 \hline
\rule[0.6em]{0pt}{0.5em} \textcolor{red}{5}  & \textcolor{red}{0.540}  & \textcolor{red}{556.693} &   \textcolor{red}{394.831}  &   \textcolor{red}{300.460}  &    \textcolor{red}{87.398}    \\
 \hline
\rule[0.6em]{0pt}{0.5em}     6  &  0.541  & 548.265 &   394.011  &   299.748  &    87.594    \\
\hline
\end{tabular}
\caption{Coupling constants of the linear $\so$ model, RMF2, that reproduce the saturation density, the binding energy per nucleon at saturation and the asymmetry energy coefficient, but not the compression modulus and the nucleon effective mass.}
\label{tab:couplingsRMF2}
\end{table}
Thus, we clearly confirm here the results obtained by the standard Walecka model~\cite{ref:Walecka}.
%
%
%
\section{Conclusions}
\label{sec:Conclusions}
From all these results, we can conclude that it is helpful~\cite{ref:Lalazissis, ref:ToddRutel}, but not necessary, to introduce the four-body self-interaction into the linear $\so$ model, as demonstrated in the present work. As long as the empirical values of the nucleon effective mass and the compression modulus cannot be estimated with high accuracy, a three-body self-interaction would be sufficient (and of course necessary), so that we can prevent getting a negative coupling constant. Briefly, $G_4$ almost always plays the role of a \textbf{negative fine-tuning parameter} of the nonlinear $\so$ model for bulk nuclear matter.\\

Furthermore, we expect that the same conclusions can be reached for finite nuclei, because the Lagrangian densities are not too different~(see, for example, Ref.\cite{ref:Lalazissis}); that is, the only difference is that we have to include the coupling of protons to the electromagnetic field and a derivative surface term. We point out that the parametrization of the $\so$ model was also made by Glendenning~\cite{ref:Glendenning} for \ism\ ($\eta=0$) and also found that the quartic self-coupling constant is negative. It is now, in principle, possible to determine the coupling constants of the $\so$ model analytically for asymmetric nuclear matter ($0<\eta<1$), because we have now an analytic expression for the compression modulus given in Eq.~\eqref{eq:Ketaeq}. This calculation makes the effective model especially well defined in terms of the empirical properties and the effect of the proton-neutron asymmetry on the values of the coupling constants. This study would be most welcome.\\

In summary, we have discussed the empirical properties of bulk nuclear matter obtained from modern scattering experiments. We have calculated the EoS for infinite nuclear matter within the linear and nonlinear $\so$ model at finite temperature. Afterward, our attention is focused on the zero-temperature EoS. Within the nonlinear model, we have presented an analytical expression for the compression modulus and investigated the EoS at all nuclear densities and different proton-neutron asymmetry $\eta$. We have found that nuclear matter ceases to saturate at $\eta$ slightly larger than $0.8$.\\
In Sec.~\ref{sec:param}, we have developed an analytical method to determine the strong coupling constants from the EoS for \ism, which allow us to reproduce all the saturation properties. This method has allowed us to interpret the quartic self-coupling constant as a negative, or positive and very large fine-tuning parameter of the nonlinear $\so$ model. We have clearly confirmed~\cite{ref:Walecka} that the linear model cannot reproduce the nucleon effective mass and the compression modulus, so a nonlinear extension introducing self-coupling terms of the scalar meson field seems to be necessary. However, owing to this analytical method, we have shown that it is possible (a) to investigate the EoS in terms of $\nb$ and $\eta$ and (b) to reproduce all the five saturation properties without the quartic self-coupling constant. We have thus concluded that the latter is \underline{\textbf{not}} necessary in the $\so$ model.
%
%
%
%
\begin{acknowledgments}
This report is an extension of the first part of my diploma thesis work, which I defended on the 18th of April  2008 at the University of Rostock in Germany. It is a great pleasure to thank my former supervisors, Professor Dr. Gerd R\"opke and Professor Dr. David Blaschke for their helpful discussions.
\end{acknowledgments}
%
%
%
%
%
%
%

%
\end{document}